\documentclass[12pt]{article}
\usepackage{amsfonts,amssymb}
\usepackage[dvips]{graphics}
\usepackage{feynmf}

\newcommand{\ie}{i.e.}
\newcommand{\eg}{e.g.}

\newcommand{\be}{\begin{equation}}
\newcommand{\ee}{\end{equation}}
\newcommand{\bea}{\begin{eqnarray}}
\newcommand{\eea}{\end{eqnarray}}
\newcommand{\ba}{\begin{array}}
\newcommand{\ea}{\end{array}}
\newcommand{\bean}{\begin{eqnarray*}}
\newcommand{\eean}{\end{eqnarray*}}

\newcommand{\Cset}{{\mathbb C}}

\newcommand{\sA}{{\mathcal A}}

\newcommand{\sL}{{\mathcal L}}

\newcommand{\bra}[1]{\langle \, #1 \, |}
\newcommand{\ket}[1]{| \, #1 \, \rangle}
\newcommand{\rep}{\stackrel{\bullet}{=}}
\newcommand{\eva}{\stackrel{I}{\longrightarrow}}
\newcommand{\pv}[1]{\parbox{3.5mm}{\vspace{-0mm}#1}}
\newcommand{\ps}[1]{\parbox{4mm}{\vspace{-1mm}#1}}
\newcommand{\pu}[1]{\parbox{5mm}{\vspace{-1mm}#1\vspace{-1mm}}}

\newcommand{\fr}[2]{{\scriptstyle {#1 \over #2}}}

\begin{document}

\hfill DMA-01-xx 

\hfill  hep-th/0105315

\begin{center}
{\LARGE
Representations of the Renormalization Group
\\ [2mm] as Matrix Lie Algebra}
\ \\ \ \\
{\large M. Berg} \footnote{This work was
    partially performed while
   visiting DMA, Ecole Normale Sup{\'e}rieure.}
\footnote
    {Email contact: mberg@physics.utexas.edu} \\
    {\it Center for Relativity} \ \\[-1mm] 
    {\it University
    of Texas at Austin, USA} 
\ \\ \ \\
{\large P. Cartier} \\ 
   {\it Departement de Math{\'e}matiques et
        Applications }  \\[-1mm]
   {\it Ecole Normale Sup{\'e}rieure, Paris, France} \\ \ \\
\ \\ \ \\
{\bf Abstract}
\end{center}

\noindent
Renormalization is cast in the form of a Lie
algebra of infinite triangular matrices. 
By exponentiation, these matrices generate
counterterms for Feynman diagrams with subdivergences. 
As representations of an insertion operator,
the matrices are related to
the Connes-Kreimer Lie algebra.
In fact, the right-symmetric nonassociative
algebra of the Connes-Kreimer insertion product is equivalent
to an ``Ihara bracket'' in the matrix Lie algebra. 
We check our results in a three-loop example in scalar field theory. 
Apart from possible applications in high-precision phenomenology,
we give a few ideas about possible applications in 
noncommutative geometry and functional
integration.

\ \\ 

\section{Introduction}

It has been known for several years now that renormalization is
governed by certain algebras known as Hopf algebras 
(for a review, see Kreimer \cite{K4}). Unfortunately,
since Hopf algebra is largely unfamiliar to physicists\footnote{except in
completely different contexts, as symmetries of some nonlinear sigma
models in low-dimensional supergravity. 
A basic Hopf algebra reference is \cite{Kassel}.}, this intriguing fact 
has not been widely appreciated.
Here, we make an attempt to help remedy this situation,
by focusing instead on an equivalent Lie algebra.
The existence of such a Lie algebra
is guaranteed by the famous Milnor-Moore theorem 
(see e.g.\ \cite{Chapoton}), 
a fact that was exploited by Connes and Kreimer already in 
\cite{CK1}.
This theorem 
says, loosely speaking, that any graded Hopf algebra can be 
written in terms of a Lie algebra. 
The mere fact that Lie algebra is much more familiar to physicists
than Hopf algebra makes this direction useful to explore. 
In addition, there are 
certain computational advantages that may prove important;
the Lie algebra can be expressed in terms of (in general infinite) matrices,
so renormalization becomes a matter of matrix algebra.
The matrix Lie algebra is the subject of this work.
We explore the application to renormalization
and give a number of examples.

Via the homomorphism between the Hopf algebra of renormalization
and the Hopf algebra of coordinates on the group of formal
diffeomorphisms in noncommutative geometry \cite{CK2,CM},
we also expect to be able to apply the matrix framework in noncommutative
geometry terms, but this direction is postponed to future work.
(Reasons why noncommutative geometry could be relevant to physics
abound \cite{ChC,CDS,MGV,SW}.)
There is also another,
more speculative way in which this matrix 
Lie algebra may be applied, explained in the conclusion,
section \ref{sec:div}. First, let us step back and review some
existing directions in the literature.
Partly through indications of recommended references,
we will attempt to maintain some accessibility
for mathematicians who
are not completely familiar with physics jargon.

\section{Background}

Perturbative renormalization
was regarded by most physicists
as finished with a theorem by 
Bogoliubov, Parasiuk, Hepp, and Zimmermann (the BPHZ theorem), 
refined by Zimmermann in 1970. Original references for this theorem 
are \cite{BPHZ}, and a clear textbook version
is given in \cite{Collins}. 
Already hidden within the BPHZ theorem, however, 
was algebraic 
structure belonging to branches of mathematics that were probably 
largely unknown to physicists in the 1970s.
The Hopf algebra of renormalization was
first displayed by Kreimer in 1997, partially as a result of his excursions
into knot theory \cite{K1}. 
In 1999, Broadhurst and Kreimer performed a Hopf-algebraic
show of strength by computing contributions to anomalous
dimensions in Yukawa theory to 30 loops \cite{BK}\footnote{This was later
improved further \cite{BK3}.}. 

On a practical level, phenomenologists have been opening their eyes to
Hopf algebra as a potential time-saver and 
organizing principle in massive computations, such as the 
5-loop computation of the beta function in quantum chromodynamics (QCD). 
Such high-precision analytical calculations are needed 
for comparisons to numerical 
calulations in lattice QCD, 
since the energy scales accessible in these lattice computations 
(and in many experiments) are often so low that perturbative
QCD is only starting to become a valid approximation
(see e.g.\ \cite{BBL,CS}).
Even when one is not seeking to push calculations to such heights of
precision, savings in computer time through better algorithms 
could be 
welcome for routine calculations as well.

On the more mathematical side, a series of articles by Connes and
Kreimer \cite{CK1,CK2} 
took the Hopf algebra of renormalization 
through some formal developments, amongst other the
identification of the Lie algebra of the dual of the aforementioned
Hopf algebra,
through the Milnor-Moore theorem (see also
\cite{Chapoton} for related 
mathematical developments). This has later been phrased as a 
study of {\it operads}\footnote{For instance, 
a simple example of an operad is a collection of maps from 
all tensor products of an algebra $\sA$ into $\sA$ itself, such as
$\sA \otimes \sA \otimes \sA \rightarrow \sA$, with some compatibility
requirements. See e.g.\ \cite{Kontsevich} for a precise definition.},
which we will not go into here, but which can be seen as
complementary to our discussion.
Very recently, some geometric 
aspects were explored in \cite{CQRV}.

We believe that representations of the dual
Lie algebra can be useful in their own right, so this work is about one such 
representation. The Lie bracket is simply given by inserting one graph
into another, and subtracting the opposite insertion. 
The process of insertion may be represented by
the multiplication of (in general infinite) matrices, hence matrix
Lie algebra.

Although we give examples in scalar field theory, the
algebra representation on graphs is independent of the specific nature of 
the action, except that we assume the interactions 
are quartic. We could equally well consider 
fermions, vector fields, or higher-spin fields with quartic
interactions. But surely 
Ward identities, ghosts and so on make gauge theory computations very
different from those of scalar field theory? 
This question can be answered on two levels. First, the
distinction between Feynman graphs and their values.
In performing field theory 
computations, one routinely identifies graphs and
their values. In the Connes-Kreimer approach one
explicitly separates symmetries of the algebra of Feynman graphs from
symmetries of Feynman integrals; the integrals are thought of as
elements of the dual
of the space of graphs. The dual has symmetries of its own,
``quasi-shuffles'' \cite{K3}, obviously related but not identical to
symmetries on graphs. 
The algebraic approach can help 
with  computing Feynman integrals,
but its greatest strengths so far has been in the
organization of counterterms, 
providing algebraic relations between seemingly unrelated quantities. 
The organization of counterterms
is an essentially graph-theoretic problem, since
the hierarchy of subdivergences
can be read off from the topology of a graph without knowing
precisely what fields are involved.
In the  extreme case of ``rainbow'' or ``ladder'' diagrams only, the Feynman
integrals themselves are easily iterated to any loop order
(as in \cite{BK}), but the 
BPHZ subtractions quickly become a combinatorial mess if
the algebraic structure is ignored.

The attitude of separating
the combinatorial and analytical problems 
(space of graphs and its dual) is also
useful for computer implementation.
Let us take one example:
the Mathematica package
{\it FeynArts} \cite{FeynArts}, which
can automatically compute
amplitudes up to one loop in any renormalizable theory,
and using additional software, up to two loops \cite{Heinemeyer}.
This particular software package calculates amplitudes
by first writing down scalar graphs with
appropriate vertices, then generates counterterm graphs,
then at a later stage lets the user specify what fields are
involved, either gauge fields, spinors or scalars. 
Since the number of graphs grows very quickly with loop order, it will
become important to take advantage of algebraic structure if
calculations at very high precision are to be feasible in a
reasonable amount of computer time.

Apart from the direct 
application of the matrix Lie algebra to precision computations, 
there are other possible indirect  
applications. Such applications may 
be found in noncommutative geometry through the result of \cite{CK2}. 
A currently more speculative
application is to functional integration. 
These directions are not explored in detail in
this work, but see section \ref{sec:div}.

\section{Graph Summary}
We consider $\phi^4$ theory in four spacetime dimensions for ease of
exposition (see the previous section for comments on adaption to other
theories). In $\phi^4$ theory in four dimensions, 
we consider graphs with a quartic interaction 
vertex\footnote{\ie\  
the interaction vertex is four-valent.} and
two and four external legs (see section \ref{sec:Wilson} for a
reminder of why we do not need e.g. six external legs in this 
theory in four dimensions).
All such graphs with up to three loops are summarized 
in table \ref{tab:graphs}\footnote{These graphs were drawn using
FeynMF \cite{FeynMF}.}. In this table, 
``crossing'' refers to graphs related to previously given graphs by
crossing symmetry, such as the two graphs in parenthesis for $L=1$,
$E=4$; they are simply found by turning the first graph on its
side (second graph), and then crossing the two lower external legs
(third graph).\footnote{The reader unfamiliar with 
Feynman diagrams may want to
note that the third graph in $L=1$, $E=4$ still has only two 
vertices: the two lines crossing below the loop are going one above
the other.}
On a computer, graphs can be stored as lists. 
One example of such lists, adapted to the context of Feynman graphs,
is given at the end of section \ref{sec:grafting}.

\begin{fmffile}{bcgraph}

\begin{table}
\begin{center}
\begin{tabular}{|l|l|cccccccc|} \hline 
$L$ & $E$ & \multicolumn{7}{c}{Graphs}& \\  \hline

0 & 2 &
\begin{fmfgraph*}(10,10)
  \fmfkeep{s0}
  \fmfpen{.8thin}
  \fmfleft{i}
  \fmfright{o}
  \fmf{plain}{i,o}
\end{fmfgraph*} & & & & & & & \\ 

0 & 4 & 
\begin{fmfgraph*}(10,10)
  \fmfkeep{v0}
  \fmfpen{.8thin}
  \fmfleft{i1,i2}
  \fmfright{o1,o2}
  \fmf{plain}{i1,o2}
  \fmf{plain}{i2,o1}
\end{fmfgraph*} & & & & & & & \\  

1 & 2 &

\begin{fmfgraph*}(10,10)
  \fmfkeep{s1}
  \fmfpen{.8thin}
  \fmfbottom{i,o}
  \fmf{plain,tension=0.5}{i,v,v,o}
\end{fmfgraph*} & & & & & & & \\

1 & 4 & 

\begin{fmfgraph*}(10,10)
  \fmfkeep{v1}
  \fmfpen{.8thin}
  \fmfleft{i1,i2}
  \fmfright{o1,o2}
  \fmfforce{.5w,.2h}{v1}
  \fmfforce{.5w,.8h}{v2}
  \fmf{plain,left=.8,tension=0.4}{v1,v2,v1}
  \fmf{plain}{i1,v1,o1}
  \fmf{plain}{i2,v2,o2}
\end{fmfgraph*} &

$\left(

\begin{fmfgraph*}(12,8)
  \fmfkeep{v2}
  \fmfpen{.8thin}
  \fmfleft{i1,i2}
  \fmfright{o1,o2}
  \fmf{plain,left=.8,tension=0.3}{v1,v2,v1}
  \fmf{plain}{i1,v1,i2}
  \fmf{plain}{o1,v2,o2}
\end{fmfgraph*} \right.$ &
$\left.
\begin{fmfgraph*}(12,8)
  \fmfkeep{v3}
  \fmfpen{.8thin}
  \fmfleft{i1,i2}
  \fmfright{o1,o2}
  \fmfforce{.2w,.5h}{v1}
  \fmfforce{.8w,.5h}{v2}
  \fmf{plain,left=.6,tension=.3}{v1,v2,v1}
  \fmf{plain,right=.3}{i1,v2}
  \fmf{plain,right=.4,rubout}{v1,o1}
  \fmf{plain}{v1,i2}
  \fmf{plain}{v2,o2}
\end{fmfgraph*} \right)$
 & & & & & \\

2 & 2 &

\begin{fmfgraph*}(12,10)
  \fmfkeep{s2}
  \fmfpen{.8thin}
  \fmfbottom{i,o}
  \fmf{plain}{i,v1,o}
  \fmffreeze
  \fmfforce{.5w,.5h}{v2}
  \fmf{plain,left=.8,tension=0.2}{v1,v2,v1}
  \fmfforce{.5w,.8h}{v3}
  \fmf{plain,left=.9,tension=0.1}{v2,v3,v2}
\end{fmfgraph*} &

\begin{fmfgraph*}(10,10)
  \fmfkeep{s3}
  \fmfpen{.8thin}
  \fmfleft{i}
  \fmfright{o}
  \fmf{plain}{i,v1,v2,o}
  \fmf{plain,left,tension=0.1}{v1,v2,v1}
  \fmfforce{.2w,.5h}{v1}
  \fmfforce{.8w,.5h}{v2}
\end{fmfgraph*} & & & & & & \\

2 & 4  &

\begin{fmfgraph*}(10,10)
  \fmfkeep{v4}
  \fmfpen{.8thin}
  \fmfleft{i1,i2}
  \fmfright{o1,o2}
  \fmfforce{.5w,.2h}{v1}
  \fmfforce{.5w,.5h}{v2}
  \fmfforce{.5w,.8h}{v3}
  \fmf{plain,left=1,tension=0.4}{v1,v2,v1}
  \fmf{plain,left=1,tension=0.4}{v2,v3,v2}
  \fmf{plain}{i1,v1,o1}
  \fmf{plain}{i2,v3,o2}
\end{fmfgraph*} &

\begin{fmfgraph*}(10,10)
  \fmfkeep{v5}
  \fmfpen{.8thin}
  \fmfleft{i1,i2}
  \fmfright{o1,o2}
  \fmf{plain,right=.3,tension=0.4}{v1,v3,v2}
  \fmf{plain,left=.6,tension=0.1}{v1,v2,v1}
  \fmf{plain}{i1,v3,o1}
  \fmf{plain}{i2,v1}
  \fmf{plain}{v2,o2}
\end{fmfgraph*} &

\begin{fmfgraph*}(10,10)
  \fmfkeep{v6}
  \fmfpen{.8thin}
  \fmfleft{i1,i2}
  \fmfright{o1,o2}
  \fmf{plain,left=.3,tension=0.4}{v1,v3,v2}
  \fmf{plain,left=.6,tension=0.1}{v1,v2,v1}
  \fmf{plain}{i2,v3,o2}
  \fmf{plain}{i1,v1}
  \fmf{plain}{v2,o1}
\end{fmfgraph*} &

\begin{fmfgraph*}(10,10)
  \fmfkeep{v7}
  \fmfpen{.8thin}
  \fmfleft{i1,i2}
  \fmfright{o1,o2}
  \fmfforce{.5w,.2h}{v1}
  \fmfforce{.5w,.8h}{v2}
  \fmfforce{.3w,.5h}{v3}
  \fmfforce{.7w,.5h}{v4}
  \fmfforce{0,.5h}{v5}
  \fmf{plain,left=.4,tension=0.4}{v1,v3,v2}
  \fmf{plain,left=.4,tension=0.4}{v2,v4,v1}
  \fmf{plain,left=.8,tension=0.2}{v3,v5,v3}
  \fmf{plain}{i1,v1,o1}
  \fmf{plain}{i2,v2,o2}
\end{fmfgraph*} 
& 
\multicolumn{3}{l}{+ crossing} & \\ 

3 & 2 &
\begin{fmfgraph*}(13,13)
  \fmfkeep{s4}
  \fmfpen{.8thin}
  \fmfbottom{i,o}
  \fmf{plain}{i,v1,o}
  \fmffreeze
  \fmfforce{.4w,.7h}{v2}
  \fmfforce{.6w,.7h}{v3}  
  \fmfforce{.25w,h}{v4}  
  \fmfforce{.75w,h}{v5}  
  \fmf{plain,left=.5}{v1,v2}
  \fmf{plain,left=.2}{v2,v3}
  \fmf{plain,left=.5}{v3,v1}
  \fmf{plain,left=.8}{v2,v4,v2}
  \fmf{plain,left=.8}{v3,v5,v3}
\end{fmfgraph*} &

\begin{fmfgraph*}(13,13)
  \fmfkeep{s5}
  \fmfpen{.8thin}
  \fmfbottom{i,o}
  \fmf{plain}{i,v1,o}
  \fmffreeze
  \fmfforce{.5w,.45h}{v2}
  \fmfforce{.5w,.75h}{v3}  
  \fmfforce{.5w,h}{v4}  
  \fmf{plain,left=.8}{v1,v2,v1}
  \fmf{plain,left=.8}{v2,v3,v2}
  \fmf{plain,left=.8}{v3,v4,v3}
\end{fmfgraph*} &

\begin{fmfgraph*}(13,13)
  \fmfkeep{s6}
  \fmfpen{.8thin}
  \fmfleft{i}
  \fmfright{o}
  \fmfforce{.15w,.5h}{v1}
  \fmfforce{.5w,.5h}{v2}  
  \fmfforce{.85w,.5h}{v3}  
  \fmf{plain,left=.6}{v1,v2,v1}
  \fmf{plain,left=.6}{v2,v3,v2}
  \fmf{plain,right}{v1,v3}
  \fmf{plain}{i,v1}
  \fmf{plain}{v3,o}
\end{fmfgraph*} &

\begin{fmfgraph*}(13,13)
  \fmfkeep{s7}
  \fmfpen{.8thin}
  \fmfbottom{i,o}
  \fmfforce{.5w,.1h}{v1}
  \fmfforce{.3w,.7h}{v2}
  \fmfforce{.7w,.7h}{v3}
  \fmf{plain}{v2,v3}
  \fmf{plain,left}{v2,v3,v2}
  \fmf{plain,right=.3}{v2,v1,v3}
  \fmf{plain}{i,v1,o}
\end{fmfgraph*}  &

\begin{fmfgraph*}(10,10)
  \fmfkeep{s8}
  \fmfpen{.8thin}
  \fmfleft{i}
  \fmfright{o}
  \fmfforce{.2w,.5h}{v1}
  \fmfforce{.8w,.5h}{v2}
  \fmfforce{.5w,.7h}{v3}
  \fmfforce{.5w,h}{v4}
  \fmf{plain}{i,v1,v2,o}
  \fmf{plain,right}{v1,v2}
  \fmf{plain,left=.4}{v1,v3,v2}
  \fmf{plain,left=.8}{v3,v4,v3}
\end{fmfgraph*}& & &  \\



3 & 4 &

\begin{fmfgraph*}(15,15)
  \fmfkeep{u1}
  \fmfpen{.8thin}
  \fmfleft{i1,i2}
  \fmfright{o1,o2}
  \fmfforce{.3w,.7h}{v1}
  \fmfforce{.7w,.7h}{v2}
  \fmfforce{.3w,.3h}{v3}
  \fmfforce{.7w,.3h}{v4}
  \fmfforce{.4w,.4h}{v5}
  \fmfforce{.6w,.6h}{v6}
  \fmf{plain,left=.5}{v1,v2,v4,v3,v1}
  \fmf{plain}{v1,v4}
  \fmf{plain}{v2,v6}
  \fmf{plain}{v5,v3}
  \fmf{plain}{i1,v3}
  \fmf{plain}{v4,o1}
  \fmf{plain}{i2,v1}
  \fmf{plain}{v2,o2}
\end{fmfgraph*} &

\begin{fmfgraph*}(15,15)
  \fmfkeep{u2}
  \fmfpen{.8thin}
  \fmfleft{i1,i2}
  \fmfright{o1,o2}
  \fmfforce{.3w,.7h}{v1}
  \fmfforce{.7w,.7h}{v2}
  \fmfforce{.3w,.3h}{v3}
  \fmfforce{.7w,.3h}{v4}
  \fmf{plain,left=.5}{v1,v2,v4,v3,v1}
  \fmf{plain,right=.5}{v1,v2}
  \fmf{plain,left=.5}{v3,v4}
  \fmf{plain}{i1,v3}
  \fmf{plain}{v4,o1}
  \fmf{plain}{i2,v1}
  \fmf{plain}{v2,o2}
\end{fmfgraph*} &

\begin{fmfgraph*}(15,15)
  \fmfkeep{u3}
  \fmfpen{.8thin}
  \fmfleft{i1,i2}
  \fmfright{o1,o2}
  \fmfforce{.3w,.7h}{v1}
  \fmfforce{.7w,.7h}{v2}
  \fmfforce{.3w,.3h}{v3}
  \fmfforce{.7w,.3h}{v4}
  \fmf{plain,left=.5}{v1,v2,v4,v3,v1}
  \fmf{plain,left=.5}{v1,v3}
  \fmf{plain,right=.5}{v2,v4}
  \fmf{plain,right=.3}{i1,v4}
  \fmf{plain,right=.3,rubout}{v3,o1}
  \fmf{plain}{i2,v1}
  \fmf{plain}{v2,o2}
\end{fmfgraph*} &

\begin{fmfgraph*}(15,15)
  \fmfkeep{u4}
  \fmfpen{.8thin}
  \fmfleft{i1,i2}
  \fmfright{o1,o2}
  \fmfforce{.5w,.2h}{v1}
  \fmfforce{.3w,.5h}{v2}
  \fmfforce{.7w,.5h}{v3}
  \fmfforce{.5w,.8h}{v4}
  \fmf{plain,left=.4}{v2,v3,v2}
  \fmf{plain,left=.8}{v1,v4,v1}
  \fmf{plain}{i1,v1,o1}
  \fmf{plain}{i2,v4,o2}
\end{fmfgraph*} &

\begin{fmfgraph*}(15,15)
  \fmfkeep{u5}
  \fmfpen{.8thin}
  \fmfleft{i1,i2}
  \fmfright{o1,o2}
  \fmfforce{.5w,.2h}{v1}
  \fmfforce{.5w,.4h}{v2}
  \fmfforce{.5w,.6h}{v3}
  \fmfforce{.5w,.8h}{v4}
  \fmf{plain,left=.8}{v1,v2,v1}
  \fmf{plain,left=.8}{v2,v3,v2}
  \fmf{plain,left=.8}{v3,v4,v3}
  \fmf{plain}{i1,v1,o1}
  \fmf{plain}{i2,v4,o2}
\end{fmfgraph*} &

\begin{fmfgraph*}(15,15)
  \fmfkeep{u6}
  \fmfpen{.8thin}
  \fmfleft{i1,i2}
  \fmfright{o1,o2}
  \fmfforce{.2w,.7h}{v1}
  \fmfforce{.5w,.7h}{v2}
  \fmfforce{.8w,.7h}{v3}
  \fmfforce{.5w,.2h}{v4}
  \fmf{plain,left=.6}{v1,v2,v1}
  \fmf{plain,left=.6}{v2,v3,v2}
  \fmf{plain,left=.4}{v3,v4,v1}
  \fmf{plain}{i2,v1}
  \fmf{plain}{v3,o2}
  \fmf{plain}{i1,v4,o1}
\end{fmfgraph*} &

\begin{fmfgraph*}(15,15)
  \fmfkeep{u7}
  \fmfpen{.8thin}
  \fmfleft{i2,i1}
  \fmfright{o2,o1}
  \fmfforce{.2w,.2h}{v1}
  \fmfforce{.5w,.2h}{v2}
  \fmfforce{.8w,.2h}{v3}
  \fmfforce{.5w,.7h}{v4}
  \fmf{plain,left=.6}{v1,v2,v1}
  \fmf{plain,left=.6}{v2,v3,v2}
  \fmf{plain,right=.4}{v3,v4,v1}
  \fmf{plain}{i2,v1}
  \fmf{plain}{v3,o2}
  \fmf{plain}{i1,v4,o1}
\end{fmfgraph*}   &

\begin{fmfgraph*}(15,15)
  \fmfkeep{u8}
  \fmfpen{.8thin}
  \fmfleft{i1,i2}
  \fmfright{o1,o2}
  \fmfforce{.3w,.8h}{v1}
  \fmfforce{.7w,.8h}{v2}
  \fmfforce{.5w,.5h}{v3}
  \fmfforce{.5w,.2h}{v4}
  \fmf{plain,right=.3}{v1,v3,v2}
  \fmf{plain,left=.6}{v1,v2,v1}
  \fmf{plain,left=.8}{v3,v4,v3}
  \fmf{plain}{i1,v4,o1}
  \fmf{plain}{i2,v1}
  \fmf{plain}{v2,o2}
\end{fmfgraph*}\\

& &  

\begin{fmfgraph*}(15,15)
  \fmfkeep{u9}
  \fmfpen{.8thin}
  \fmfleft{i1,i2}
  \fmfright{o1,o2}
  \fmfforce{.3w,.2h}{v1}
  \fmfforce{.7w,.2h}{v2}
  \fmfforce{.5w,.5h}{v3}
  \fmfforce{.5w,.8h}{v4}
  \fmf{plain,left=.3}{v1,v3,v2}
  \fmf{plain,left=.6}{v1,v2,v1}
  \fmf{plain,left=.8}{v3,v4,v3}
  \fmf{plain}{i2,v4,o2}
  \fmf{plain}{i1,v1}
  \fmf{plain}{v2,o1}
\end{fmfgraph*} &

\begin{fmfgraph*}(15,15)
  \fmfkeep{u10}
  \fmfpen{.8thin}
  \fmfleft{i1,i2}
  \fmfright{o1,o2}
  \fmfforce{.7w,.65h}{v1}
  \fmfforce{.3w,.8h}{v2}
  \fmfforce{.3w,.5h}{v3}
  \fmfforce{.5w,.2h}{v4}
  \fmf{plain,left=.8}{v2,v3,v2}
  \fmf{plain,right=.4}{v1,v2}
  \fmf{plain,right=.4}{v3,v1}  
  \fmf{plain,left=.3}{v1,v4}
  \fmf{plain,left=.2}{v4,v3}
  \fmf{plain}{i2,v2}
  \fmf{plain}{v1,o2}
  \fmf{plain}{i1,v4,o1}
\end{fmfgraph*} &

\begin{fmfgraph*}(15,15)
  \fmfkeep{u11}
  \fmfpen{.8thin}
  \fmfleft{i2,i1}
  \fmfright{o2,o1}
  \fmfforce{.7w,.35h}{v1}
  \fmfforce{.3w,.2h}{v2}
  \fmfforce{.3w,.5h}{v3}
  \fmfforce{.5w,.8h}{v4}
  \fmf{plain,right=.8}{v2,v3,v2}
  \fmf{plain,left=.4}{v1,v2}
  \fmf{plain,left=.4}{v3,v1}  
  \fmf{plain,right=.3}{v1,v4}
  \fmf{plain,right=.2}{v4,v3}
  \fmf{plain}{i2,v2}
  \fmf{plain}{v1,o2}
  \fmf{plain}{i1,v4,o1}
\end{fmfgraph*} &

\begin{fmfgraph*}(15,15)
  \fmfkeep{u12}
  \fmfpen{.8thin}
  \fmfleft{i1,i2}
  \fmfright{o1,o2}
  \fmfforce{.3w,.65h}{v1}
  \fmfforce{.7w,.8h}{v2}
  \fmfforce{.7w,.5h}{v3}
  \fmfforce{.5w,.2h}{v4}
  \fmf{plain,left=.8}{v2,v3,v2}
  \fmf{plain,left=.4}{v1,v2}
  \fmf{plain,left=.4}{v3,v1}  
  \fmf{plain,right=.3}{v1,v4}
  \fmf{plain,right=.2}{v4,v3}
  \fmf{plain}{i2,v1}
  \fmf{plain}{v2,o2}
  \fmf{plain}{i1,v4,o1}
\end{fmfgraph*} &

\begin{fmfgraph*}(15,15)
  \fmfkeep{u13}
  \fmfpen{.8thin}
  \fmfleft{i1,i2}
  \fmfright{o1,o2}
  \fmfforce{.3w,.35h}{v1}
  \fmfforce{.7w,.2h}{v2}
  \fmfforce{.7w,.5h}{v3}
  \fmfforce{.5w,.8h}{v4}
  \fmf{plain,right=.8}{v2,v3,v2}
  \fmf{plain,right=.4}{v1,v2}
  \fmf{plain,right=.4}{v3,v1}  
  \fmf{plain,left=.3}{v1,v4}
  \fmf{plain,left=.2}{v4,v3}
  \fmf{plain}{i1,v1}
  \fmf{plain}{v2,o1}
  \fmf{plain}{i2,v4,o2}
\end{fmfgraph*} &
  
\begin{fmfgraph*}(15,15)
  \fmfkeep{u14}
  \fmfpen{.8thin}
  \fmfleft{i1,i2}
  \fmfright{o1,o2}
  \fmfforce{.5w,.8h}{v1}
  \fmfforce{.5w,.2h}{v2}
  \fmfforce{.35w,.65h}{v3}
  \fmfforce{.35w,.35h}{v4}
  \fmf{plain,right=.2}{v1,v3,v4,v2}
  \fmf{plain,right}{v3,v4,v3}
  \fmf{plain,right=.8}{v2,v1}
  \fmf{plain}{i1,v2,o1}
  \fmf{plain}{i2,v1,o2}
\end{fmfgraph*}  &
 
\begin{fmfgraph*}(15,15)
  \fmfkeep{u15}
  \fmfpen{.8thin}
  \fmfleft{i1,i2}
  \fmfright{o1,o2}
  \fmfforce{.5w,.8h}{v1}
  \fmfforce{.5w,.2h}{v2}
  \fmfforce{.4w,.5h}{v3}
  \fmfforce{.2w,.5h}{v4}
  \fmfforce{0,.5h}{v5}
  \fmf{plain,right=.3}{v1,v3,v2}
  \fmf{plain,right=.8}{v3,v4,v3}
  \fmf{plain,right=.8}{v4,v5,v4}
  \fmf{plain,right=.5}{v2,v1}
  \fmf{plain}{i1,v2,o1}
  \fmf{plain}{i2,v1,o2}
\end{fmfgraph*} &
 
\begin{fmfgraph*}(15,15)
  \fmfkeep{u16}
  \fmfpen{.8thin}
  \fmfleft{i1,i2}
  \fmfright{o1,o2}
  \fmfforce{.5w,.8h}{v1}
  \fmfforce{.5w,.2h}{v2}
  \fmfforce{.3w,.35h}{v3}
  \fmfforce{.3w,.65h}{v4}
  \fmfforce{.1w,.35h}{v5}
  \fmfforce{.1w,.65h}{v6}
  \fmf{plain,right=.3}{v1,v4,v3,v2}
  \fmf{plain,right=.8}{v3,v5,v3}
  \fmf{plain,right=.8}{v4,v6,v4}
  \fmf{plain,right=.6}{v2,v1}
  \fmf{plain}{i1,v2,o1}
  \fmf{plain}{i2,v1,o2}
\end{fmfgraph*}\\

 & &  
 
\begin{fmfgraph*}(15,15)
  \fmfkeep{u17}
  \fmfpen{.8thin}
  \fmfleft{i1,i2}
  \fmfright{o1,o2}
  \fmfforce{.5w,.8h}{v1}
  \fmfforce{.5w,.2h}{v2}
  \fmfforce{.35w,.5h}{v3}
  \fmfforce{.65w,.5h}{v4}
  \fmfforce{.1w,.5h}{v5}
  \fmfforce{.9w,.5h}{v6}
  \fmf{plain,right=.3}{v1,v3,v2}
  \fmf{plain,left=.3}{v1,v4,v2}
  \fmf{plain,right=.8}{v3,v5,v3}
  \fmf{plain,right=.8}{v4,v6,v4}
  \fmf{plain}{i1,v2,o1}
  \fmf{plain}{i2,v1,o2}
\end{fmfgraph*} &
 
\begin{fmfgraph*}(15,15)
  \fmfkeep{u18}
  \fmfpen{.8thin}
  \fmfleft{i1,i2}
  \fmfright{o1,o2}
  \fmfforce{.5w,.8h}{v1}
  \fmfforce{.5w,.5h}{v2}
  \fmfforce{.5w,.2h}{v3}
  \fmfforce{.4w,.65h}{v4}
  \fmfforce{.15w,.65h}{v5}
  \fmf{plain,right=.2}{v1,v4,v2}
  \fmf{plain,right=.8}{v4,v5,v4}
  \fmf{plain,right=.7}{v2,v3,v2}
  \fmf{plain,right=.7}{v2,v1}
  \fmf{plain}{i1,v3,o1}
  \fmf{plain}{i2,v1,o2}
\end{fmfgraph*} &
 
\begin{fmfgraph*}(15,15)
  \fmfkeep{u19}
  \fmfpen{.8thin}
  \fmfleft{i1,i2}
  \fmfright{o1,o2}
  \fmfforce{.3w,.8h}{v1}
  \fmfforce{.7w,.8h}{v2}
  \fmfforce{.3w,.5h}{v3}
  \fmfforce{.5w,.2h}{v4}
  \fmfforce{0,.5h}{v5}
  \fmf{plain,right=.2}{v1,v3,v4}
  \fmf{plain,right=.8}{v3,v5,v3}
  \fmf{plain,left=.6}{v1,v2,v1}
  \fmf{plain,right=.4}{v4,v2}
  \fmf{plain}{i1,v4,o1}
  \fmf{plain}{i2,v1}
  \fmf{plain}{v2,o2}
\end{fmfgraph*} &
 
\begin{fmfgraph*}(15,15)
  \fmfkeep{u20}
  \fmfpen{.8thin}
  \fmfleft{i1,i2}
  \fmfright{o1,o2}
  \fmfforce{.3w,.2h}{v1}
  \fmfforce{.7w,.2h}{v2}
  \fmfforce{.3w,.5h}{v3}
  \fmfforce{.5w,.8h}{v4}
  \fmfforce{0,.5h}{v5}
  \fmf{plain,left=.2}{v1,v3,v4}
  \fmf{plain,left=.8}{v3,v5,v3}
  \fmf{plain,left=.6}{v1,v2,v1}
  \fmf{plain,left=.4}{v4,v2}
  \fmf{plain}{i2,v4,o2}
  \fmf{plain}{i1,v1}
  \fmf{plain}{v2,o1}
\end{fmfgraph*} &
 
\begin{fmfgraph*}(15,15)
  \fmfkeep{u21}
  \fmfpen{.8thin}
  \fmfleft{i1,i2}
  \fmfright{o1,o2}
  \fmfforce{.3w,.6h}{v1}
  \fmfforce{.7w,.6h}{v2}
  \fmfforce{.5w,.7h}{v3}
  \fmfforce{.5w,.2h}{v4}
  \fmfforce{.5w,h}{v5}
  \fmf{plain,left=.2}{v1,v3,v2}
  \fmf{plain,right=.8}{v3,v5,v3}
  \fmf{plain,right=.6}{v1,v2}
  \fmf{plain,right=.4}{v1,v4,v2}
  \fmf{plain}{i1,v4,o1}
  \fmf{plain}{i2,v1}
  \fmf{plain}{v2,o2}
\end{fmfgraph*} &
 
\begin{fmfgraph*}(15,15)
  \fmfkeep{u22}
  \fmfpen{.8thin}
  \fmfleft{i1,i2}
  \fmfright{o1,o2}
  \fmfforce{.3w,.4h}{v1}
  \fmfforce{.7w,.4h}{v2}
  \fmfforce{.5w,.3h}{v3}
  \fmfforce{.5w,.8h}{v4}
  \fmfforce{.5w,0}{v5}
  \fmf{plain,right=.2}{v1,v3,v2}
  \fmf{plain,right=.8}{v3,v5,v3}
  \fmf{plain,left=.6}{v1,v2}
  \fmf{plain,left=.4}{v1,v4,v2}
  \fmf{plain}{i2,v4,o2}
  \fmf{plain}{i1,v1}
  \fmf{plain}{v2,o1}
\end{fmfgraph*} & + crossing \hspace{-1cm} &   \\ \hline 
\end{tabular}
\caption{Graphs with $L$ loops and $E$ external legs.}
\label{tab:graphs}
\end{center}
\end{table}

\section{The Grafting Operator}
\label{sec:grafting}
\begin{figure}[h]
  \begin{center}
     \resizebox{4cm}{!}{\includegraphics{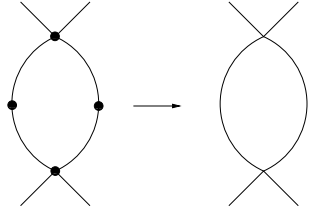}}
   \end{center}
\caption{
Insertion points are implicit.}
\label{insertion}
\end{figure}

The Lie algebra of graphs comes from inserting (grafting) one Feynman
graph into another. 
First a remark about terminology. 
In standard Feynman graphs, each internal line represents a
propagator. 
We will only work
with amputated graphs, where the external legs do not represent
propagators. 
Now, in a theory with quartic 
interactions, one can consider
inserting graphs with either two external legs (into an internal line)
or four external legs (into an interaction vertex).
Thus, the middle of an internal line and the interaction vertex
itself are eligible as {\it insertion points}.
If desired, we could
represent each insertion point by a dot, as in 
the graph on the left in fig.\ \ref{insertion}. 
As in standard usage---the graph on the right  in 
fig.\ \ref{insertion}---we leave those dots implicit.
We alert the reader that the authors of
\cite{CK1} call
the dots on lines ``two-point vertices''.
To avoid confusion with some of the physics terminology,
we prefer to talk about insertion points instead of 
two-point vertices.

Here is a precise definition of the insertion, or grafting,
operator. The practically-minded reader can refer to the insertion 
tables in appendix 
\ref{app:E}
to get a quick idea of the action of this operator.
Enumerate the insertion points
of a graph $t$ by $p_{i}(t)$, the total number of insertion points
of $t$ by 
$N(t)$ and their valences (number of ``prongs'', \ie\ lines
sticking out from the vertex) by $v(p_{i}(t))$.
Here, the valence is always either 2 or 4.
If $t_1$ and $t_2$ are graphs, and the number $v(t_2)$ of external legs
of $t_2$ is equal to the valence $v(p_i(t_1))$ of the insertion point
$p_i(t_1)$ of $t_1$, then we define
$s_{t_2}^{p_i(t_1)}$ to be insertion of the graph $t_2$ at 
$p_i(t_1)$ by summing over all permutations of the $v(p_{i}(t_1))$
propagators to be joined. Then the grafting operator $s_{t_2}$ is
\be
s_{t_2} t_1 = 
{1 \over v(t_2)!} \sum_{i=1}^{N(t_1)} s_{t_2}^{p_i(t_1)} t_1 \; .
\label{defs}
\ee
The total number of graphs  ${\mathcal N}$
(including multiplicities of the same
graph) created by insertion of $t_2$ into $t_1$ is
\[
{\mathcal N}=
\sum_{i=1}^{N(t_1)} \delta_{v(p_{i}(t_1)), v(p_{i}(t_2))}
v(p_{i}(t_1))!
\]
We call $t_1$ the ``object graph'' and $t_2$ the ``insertion graph''.  
Often, we will use parenthesis to delineate the insertion graph:
$s(t_2)t_1$.
Finally, we define $s(t)$ to be linear: 
\[
s(a t_1 + b t_2) = a s(t_1) + b s (t_2) 
\qquad a, b \in \Cset \; .
\]
A remark about normalization: the normalization in (\ref{defs}) 
is a mean over insertions.
It gives $s_t 1 = t$, where
$1$ represents tree-level graphs, with either 2 or 4 external legs. 
(When one needs to make a distinction, one can use $1_2$ and $1_4$.)
On the other hand, we have $s_1 t \neq t$, in fact
\[
s_1 \, t = N(t)\, t = (3L(t)\pm 1)\, t \qquad (t\neq 1)
\]
where $L(t)$ is the number of loops of $t$ and the
upper (lower) sign refers to graphs with 4 (2) external legs,
respectively. 

We can assign a grading (degree) to the algebra by loop number
$L$, since then obviously
\[
\deg s_{t_2} t_1 = \deg t_2 +  \deg t_1 \; , 
\]
so within the set of graphs with a fixed number of external legs,
$s_1$ acts as our grading operator. Being the only $s_t$ which is
diagonal, we will exclude $s_1$ from the algebra, but it can be adjoined
whenever needed as in \cite{CK2}.

As an alternative to this construction of the grafting operator, one 
could consider a restricted grafting operator, 
where permutations of the legs of the insertion
which lead to the same topologies are represented by separate operators.
In this work, we concentrate on exploring the full operator.

\subsection{Nested divergences}
It is of particular interest to know when a graph has nested divergences,
i.e.\ when two divergent loops share the same propagator.
For any graph, two lines are said to be indistinguishable
if they are connected to the same vertex,
and distinguishable if they are not.  Whenever we insert
a (loop) graph with four external legs 
at a certain vertex and two distinguishable external
prongs of the insertion graph are connected to indistinguishable legs of 
another vertex, we create nested divergences.
Of course, any nested divergences already present
in the insertion graph are preserved.

\begin{figure}[h]
  \begin{center}
     \resizebox{5cm}{!}{\includegraphics{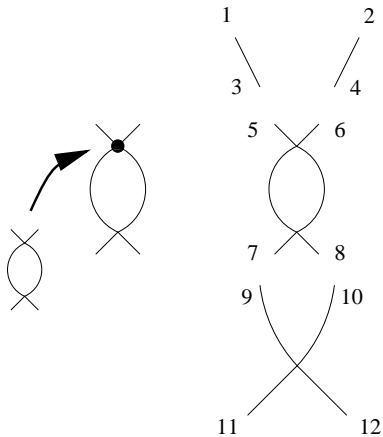}}
   \end{center}
\caption{
The creation of overlapping divergences}
\label{overlap}
\end{figure}

Here is an example. Consider the insertion in fig. \ref{overlap}.
It is an insertion of \pv{\fmfreuse{v1}} into 
the upper vertex of that same graph. 
With the above use of language, 5 and 6 are equivalent,
and so are 9 and 10, but not e.g.\ 5 and 7. 
When we join 3-5, 4-6, 7-9, 8-10, we create
\pv{\fmfreuse{v4}}, which only has two disjoint divergent loops. 
But when we join
3-5, 4-8, 7-9, 6-10, we create
\pv{\fmfreuse{v5}}, which has nested divergences.

As a side remark,
figure \ref{overlap} also lets us recall how 
\pu{\fmfreuse{v4}} is easily represented in a format amenable to
computer processing. The bottom vertex can be stored as
$[9,10,11,12]$, and
the propagator going between 5 and 7 carrying loop momentum $p$ can
be stored as $[5,7,p]$. This way, the list of vertices contains three
sublists, and the list of propagators contains four sublists, which
together completely specify the graph.
This is also how graphs are stored in {\it FeynArts} \cite{FeynArts}.
Of course, there are many other equivalent representations, 
for instance 
using the relation between rooted trees and parenthesized words 
\cite{BK2}.
\label{page:lists}

\section{Lie Algebra}
\label{sec:lie}
We want to show that the grafting operators $s_t$ form a Lie algebra
${\mathcal L}$.
To this end, we first show a certain operator identity.
Indeed for graphs $t_1$, $t_2$, define
\be
[t_1,t_2] = s(t_1)t_2 - s(t_2)t_1
\ee
then we shall show that the commutator
of the operators $s(t_1)$, $s(t_2)$ is equal to
$s([t_1,t_2])$:
\be
[s(t_1), s(t_2)] = s([t_1,t_2])   \; , 
\label{eq:ihara}
\ee
as an operator identity acting on graphs.
This identity is analogous to the Ihara bracket of the Magnus group,
familiar from number theory (see \eg\ \cite{R}).

\begin{figure}[h]
  \begin{center}
     \resizebox{11cm}{!}{\includegraphics{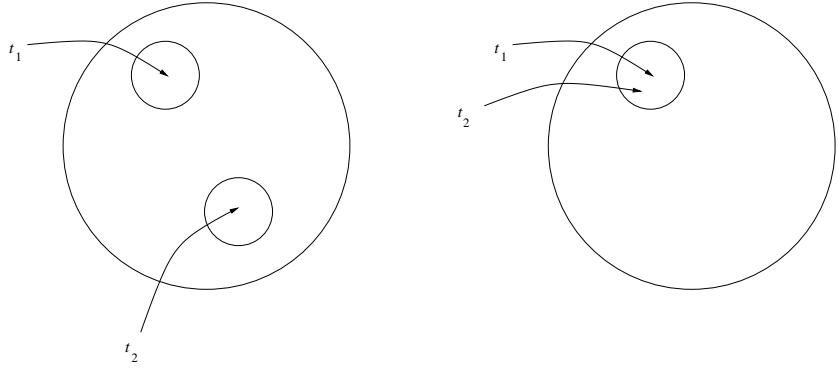}}
   \end{center}
\caption{
Difference between mutually nonlocal (a) and mutually 
local insertions (b).}
\label{ihara}
\end{figure}

Here is a sketch of the proof of (\ref{eq:ihara}).
Consider two insertions into a graph $t_3$, 
$s^{p_i(t_3)}_{t_1}$ and $s^{p_j(t_3)}_{t_2}$, 
before summing over insertion points $p_i(t_3)$ and $p_j(t_3)$. 
Let us separate insertions into ``mutually 
local'' (if $p_i(t_3)=p_j(t_3)$) and ``mutually nonlocal''
(otherwise).
This is illustrated in figure \ref{ihara}. 
The ``mutually nonlocal'' insertions clearly commute,
since they modify different points,
so these contributions to $s_{t_1}$ and $s_{t_2}$ cancel in the bracket.
We are left with the ``local'' insertions, i.e.\ 
\[
s_{t_1}t_2 - s_{t_2}t_1 =: [t_1,t_2]
\]
This concludes the outline of the proof. 
Using the correspondence in section \ref{sec:star},
a detailed proof will appear in \cite{Cartier}.

We remark that the identity trivially holds acting on $1$:
\[
[s(t_1),s(t_2)]1 = s(t_1)s(t_2)1-s(t_2)s(t_1)1
= s(t_1)t_2-s(t_2)t_1 = [t_1,t_2] \; .
\]

It is easy to check that 
the bracket $[t_1,t_2]=s(t_1)t_2-s(t_2)t_1$ 
satisfies the Jacobi identity.
By writing down three copies of the Ihara bracket identity
(\ref{eq:ihara}) and adding
them we find by linearity
\[
s([t_1,t_2])t_3 + s(t_3)[t_2,t_1] + \mbox{cyclic} = 0
\]
or
\[
[t_1,[t_2,t_3]] + \mbox{cyclic} = 0 \; .
\]

The Lie algebra is graded by loop order:
${\mathcal L}={\mathcal L}^{(1)}\oplus {\mathcal L}^{(2)}
\oplus {\mathcal L}^{(3)} \oplus \ldots $.
and
\[
[{\mathcal L}^{(m)}, {\mathcal L}^{(n)}] \subset 
 {\mathcal L}^{(m+n)} \; .
\]

Let us see what the Lie bracket
looks like explicitly in our field theory
with four-vertices. To make the notation more economic,
we suppress any graph which is related by crossing symmetry
to one that is already included. That is, since \pv{\fmfreuse{v1}},
\pv{\fmfreuse{v2}} and \pv{\fmfreuse{v3}} (see table \ref{tab:graphs})
all represent the same function of the different Mandelstam 
variables\footnote{Here $s=(p_1+p_2)^2$, $t=(p_1'-p_1)^2$ 
and $u=(p_2'-p_1)^2$, where $p_1$, $p_2$ are incoming momenta,
and $p_1'$, $p_2'$ are outgoing momenta.}
$s$, $t$, and $u$,
we will only display \pv{\fmfreuse{v1}} out of those three diagrams.
We can always use a symmetric renormalization point, i.e. we can define 
the theory at some mass scale $M$, meaning $s=t=u=-M^2$ at this scale,
in which case counterterms are the same for all three diagrams
mentioned previously.

At the one-loop level, we have the following diagrams:
\ps{\fmfreuse{s0}}, \pv{\fmfreuse{v0}},
\ps{\fmfreuse{s1}} and \pv{\fmfreuse{v1}}.
Consulting the insertion tables in the appendix,
or equivalently the matrix representation which we introduce below,
we find
the commutators of tree-level diagrams with one-loop diagrams:
\bean
[{\mathcal L}^{(0)},{\mathcal L}^{(1)}] 
\subset {\mathcal L}^{(1)} : \qquad \qquad && \\ [0mm]
[ \ps{\fmfreuse{s0}}, \ps{\fmfreuse{s1}} ]&=& 
s(\ps{\fmfreuse{s0}})  \ps{\fmfreuse{s1}}
- s(\ps{\fmfreuse{s1}})  \ps{\fmfreuse{s0}} \\
&=&  2\, \ps{\fmfreuse{s1}} - \ps{\fmfreuse{s1}} \\
&=&  \ps{\fmfreuse{s1}} \\[0mm]
[ \pv{\fmfreuse{v0}}, \pv{\fmfreuse{v1}} ]&=& 
s(\pv{\fmfreuse{v0}})  \pv{\fmfreuse{v1}}
- s(\pv{\fmfreuse{v1}})  \pv{\fmfreuse{v0}}\\
&=& 2\, \pv{\fmfreuse{v1}}- \pv{\fmfreuse{v1}} \\
&=& \pv{\fmfreuse{v1}}  \; .
\eean
The only nonvanishing commutator of two one-loop graphs is
\bea
[{\mathcal L}^{(1)},{\mathcal L}^{(1)}] 
\subset {\mathcal L}^{(2)} : \qquad \qquad && \nonumber \\ [0mm]
[ \pv{\fmfreuse{v1}}, \ps{\fmfreuse{s1}} ]&=&
s(\pv{\fmfreuse{v1}})  \ps{\fmfreuse{s1}}
- s(\ps{\fmfreuse{s1}})  \pv{\fmfreuse{v1}} \nonumber \\
&=& {1 \over 3} \, \ps{\fmfreuse{s2}}+ {2 \over 3}\, \pv{\fmfreuse{s3}} \, 
- 2\, \pv{\fmfreuse{v7}} \; \eva \;  
{\mathcal O}(\hbar) \; . 
\label{eq:commut}
\eea
where by $\eva$ we mean evaluation of the graphs.
Here we have restored $\hbar$ to connect with the semiclassical
approximation; an $L$-loop graph is of order $\hbar^{L-1}$.
At this point we note that the combinatorial factors 1/3, 2/3 and so
on are {\it not} the usual symmetry factors of graphs, \ie\ the
multiplicity of operator contractions generating the same graph.

Including two-loop graphs in the algebra,
the following graphs come into play:
\pv{\fmfreuse{v4}}, \pv{\fmfreuse{v5}}, 
\pv{\fmfreuse{v6}} and \pv{\fmfreuse{v7}}.
We can now consider a commutator 
of a one-loop and a two-loop graph, which creates a 
sequence of three-loop graphs:
\bean
[{\mathcal L}^{(1)},{\mathcal L}^{(2)}] 
\subset {\mathcal L}^{(3)} : \qquad \qquad && \\ [0mm]
[ \pv{\fmfreuse{v4}}, \pv{\fmfreuse{v1}} ]
&=&
s(\pv{\fmfreuse{v4}})  \pv{\fmfreuse{v1}}
- s(\pv{\fmfreuse{v1}})  \pv{\fmfreuse{v4}} \\
&=& 
{2 \over 3} \, \pu{\fmfreuse{u5}} + {2 \over 3} \, \pu{\fmfreuse{u6}} 
+ {2 \over 3} \, \pu{\fmfreuse{u7}} \\
&&-{2 \over 3} \, \pu{\fmfreuse{u4}} -   \pu{\fmfreuse{u5}} 
-{2 \over 3} \, \pu{\fmfreuse{u8}} - {2 \over 3}
\pu{\fmfreuse{u9}} \\ [2mm]
&\eva&
\; {\mathcal O}(\hbar^2) \; ,
\eean
which may be approximated by zero if we only keep
graphs up to order $\hbar$. Thus, to this order, all the 
``new''graphs with two loops are central elements of the algebra.

In general, if we consider $L$ loops (order $\hbar^{L-1}$), 
the commutator at order $\hbar^L$ is negligible, so only 
commutators of graphs of loop orders $L_1$, $L_2$ where
$L_1+L_2 \leq L$ are nonvanishing.
The effect of going from $L$ to $L+1$ is then to add central graphs
as new generators,
which appear in lower-order commutators with one extra factor of $\hbar$.
In other words, the new graphs {\it deform} the $L$-loop algebra by $\hbar$. 

Now that we have displayed a few commutators,
here are two examples of how the Ihara bracket identity works.
Let us first consider
a one-loop/one-loop commutator, acting on the bare propagator.
\bean
s([ \pv{\fmfreuse{v1}}, \ps{\fmfreuse{s1}} ])\, \ps{\fmfreuse{s0}}
&=&
s(s(\pv{\fmfreuse{v1}})\ps{\fmfreuse{s1}})\,  \ps{\fmfreuse{s0}}-
s(s(\ps{\fmfreuse{s1}})\pv{\fmfreuse{v1}})\,  \ps{\fmfreuse{s0}} \\
&=&s\left({1 \over 3}\, \ps{\fmfreuse{s2}}+{2 \over 3} \, 
\pv{\fmfreuse{s3}} \right)
\ps{\fmfreuse{s0}} \\
&=& {1 \over 3}\, 
s( \ps{\fmfreuse{s2}})\, \ps{\fmfreuse{s0}}+{2 \over 3}\,
 s(\pv{\fmfreuse{s3}}) \, \ps{\fmfreuse{s0}} \\
&=& {1 \over 3} \, \ps{\fmfreuse{s2}} + {2 \over 3} \,  \pv{\fmfreuse{s3}}
\eean
whereas
\bean
[s(\pv{\fmfreuse{v1}}), s(\ps{\fmfreuse{s1}}) ] \, \ps{\fmfreuse{s0}} &=&
s(\pv{\fmfreuse{v1}}) s(\ps{\fmfreuse{s1}}) \,  \ps{\fmfreuse{s0}}
- s(\ps{\fmfreuse{s1}}) s(\pv{\fmfreuse{v1}})\,  \ps{\fmfreuse{s0}} \\
&=& s(\pv{\fmfreuse{v1}}) \,  \ps{\fmfreuse{s1}} \\
&=& {1 \over 3}\, \ps{\fmfreuse{s2}}+{2 \over 3} \,
\pv{\fmfreuse{s3}}\; .  \\
\eean
At this level, the identity is quite trivial.
Let us therefore also check the relation at the three-loop level:
\bea
[s(\pv{\fmfreuse{v1}}), s(\ps{\fmfreuse{s1}}) ] \, \pv{\fmfreuse{v1}} &=&
s(\pv{\fmfreuse{v1}}) s(\ps{\fmfreuse{s1}}) \,  \pv{\fmfreuse{v1}}
- s(\ps{\fmfreuse{s1}}) s(\pv{\fmfreuse{v1}})\,  \pv{\fmfreuse{v1}}
\nonumber  \\
&=& s(\pv{\fmfreuse{v1}}) (2 \, \pv{\fmfreuse{v7}})-
s(\ps{\fmfreuse{s1}}) \left({2 \over 3} \, \pv{\fmfreuse{v4}}+
{2 \over 3} \, \pv{\fmfreuse{v5}}+ {2 \over 3} \,
\pv{\fmfreuse{v6}}\right)
\nonumber \\
&=& 2 \left({2 \over 3} \,  \pu{\fmfreuse{u14}}+ {1 \over 3} \, 
\pu{\fmfreuse{u15}}
+ {2 \over 3} \, \pu{\fmfreuse{u18}}+{2 \over 3} \,  \pu{\fmfreuse{u19}} 
+ {2 \over 3} \, \pu{\fmfreuse{u20}}\right)  \\
&-& {2 \over 3} \left( 4 \, \pu{\fmfreuse{u18}} 
+ 2 \, \pu{\fmfreuse{u19}}+ 2 \, \pu{\fmfreuse{u20}}
+ 2 \, \pu{\fmfreuse{u21}}+ 2 \, \pu{\fmfreuse{u22}} \right) 
\nonumber \\ 
&=& {4 \over 3} \,  \pu{\fmfreuse{u14}}+ {2 \over 3} \,  \pu{\fmfreuse{u15}}
- {4 \over 3} \, \pu{\fmfreuse{u18}} -{4 \over 3} \,  \pu{\fmfreuse{u21}} 
- {4 \over 3} \, \pu{\fmfreuse{u22}} \nonumber
\label{eq:iharach}
\eea
but
\bean
s([\pv{\fmfreuse{v1}},\ps{\fmfreuse{s1}}]) \, \pv{\fmfreuse{v1}} &=&
s\left(
{1 \over 3}\, \ps{\fmfreuse{s2}}+ {2 \over 3}\, \pv{\fmfreuse{s3}} \, 
- 2\, \pv{\fmfreuse{v7}}\right) \, \pv{\fmfreuse{v1}} \\
&=& {1 \over 3} ( 2 \,  \pu{\fmfreuse{u15}}) + {2 \over 3}
(2 \,  \pu{\fmfreuse{u14}})
- 2 \left( {2 \over 3} \,  \pu{\fmfreuse{u18}} +{2 \over 3}
 \,  \pu{\fmfreuse{u21}} 
+{2 \over 3} \, \pu{\fmfreuse{u22}}\right) \\
&=& {4 \over 3} \,  \pu{\fmfreuse{u14}}+ {2 \over 3} \,  \pu{\fmfreuse{u15}}
- {4 \over 3} \, \pu{\fmfreuse{u18}} -{4 \over 3} \,  \pu{\fmfreuse{u21}} 
- {4 \over 3} \, \pu{\fmfreuse{u22}} \; . \\
\eean
Here we see that cancellation of graphs 
$\pu{\fmfreuse{u19}}\!$ and $\pu{\fmfreuse{u20}}\!$ was
necessary in the first bracket, which is to be expected
since these two graphs are generated by ``mutually nonlocal'' insertions,
as defined above.

To summarize, if we know the commutator of two insertions, the
identity (\ref{eq:ihara}) gives us the insertion of the commutator.
A more concrete expression of this
will be seen in section \ref{sec:matrix}.

Going back to the
question of normalization, not any 
combination of normalizations will 
preserve the Ihara bracket, of course, as can easily be seen from
the example (\ref{eq:iharach}). 
One way to preserve it is to simultaneously drop
overall normalization and multiplicities, 
but without multiplicities 
the grafting operator loses some
of its combinatorial information (it ceases to count
the number of ways a graph can be generated).
Although from a physics point of view 
there is no freedom in the
choice of normalization in the total sum of renormalized
graphs --- for given Feynman rules ---
it can be useful to consider different normalizations in these 
intermediate expressions.

\subsection{Connection with the Star Operation}
\label{sec:star}
The Ihara bracket identity can be compared to the star product
discussed by Kreimer \cite{K4}. 

Here is the 
connection between our operator $s_t$ and Kreimer's star insertion
$\star$:
\[
t_1 \star t_2 \sim s(t_2) \, t_1
\]
where the $\sim$ alerts the reader that we use a different
normalization, (\ref{defs}).
 \\ \ \\
Furthermore,
\bea
(t_1 \star t_2 ) \star t_3 &\sim& s(t_3) s(t_2) t_1 \label{eq:Kr1} \\
t_1 \star (t_2 \star t_3) &\sim& s(s(t_3)t_2) t_1 \label{eq:Kr2} \; .
\eea

The coproduct $\Delta$ of the Connes-Kreimer Hopf algebra is coassociative,
so the dual Hopf algebra should have an associative multiplication 
$\odot$. One would then 
naively expect the dual Lie algebra to have an associative multiplication 
as well. However, 
comparing (\ref{eq:Kr1}) and (\ref{eq:Kr2}), 
the star operation is not associative.
This is because the original Hopf algebra contains disconnected
graphs, but only connected graphs are generated by the star product 
(or equivalently, the grafting operator $s_t$). 
When the multiplication $\odot$ is truncated to
connected graphs, it is no longer associative.

The deviation from associativity can be measured by
the {\it associativity defect}:
\be
a(t_1\; | \; t_2,t_3) \equiv
t_1 \star (t_2 \star t_3) - (t_1 \star t_2) \star
t_3 \; .
\label{eq:assdefect}
\ee
When this is nonzero, as in any nonassociative algebra, 
it is interesting to consider the ``right-symmetric'' algebra
for which
\be
a(t_1\; | \; t_2,t_3) = a(t_1\; | \; t_3,t_2)
\label{eq:leftsym}
\ee
(a ``left-symmetric'' algebra can equivalently be considered).
This leads naturally to
an algebra of the star product \cite{K4}, which is also
common usage in differential geometry.
Writing the right-symmetric condition
(\ref{eq:leftsym}) explicitly in terms of the associativity defect
(\ref{eq:assdefect}), the algebra is specified by
\be
t_1 \star (t_2 \star t_3) - (t_1 \star t_2) \star t_3 
= t_1 \star (t_3 \star t_2) - (t_1 \star t_3) \star t_2 \; .
\label{eq:preLie}
\ee
This is known as a {\it pre-Lie} (or {\it Vinberg}) algebra.
A pre-Lie algebra yields a bracket $[t_1,t_2]=t_1 \star t_2 - t_2 
\star t_1$ that automatically satisfies the Jacobi identity, 
hence ``pre-Lie''.
In our notation, using eqs.\ (\ref{eq:Kr1}),
(\ref{eq:Kr2}), the pre-Lie algebra identity 
(\ref{eq:preLie}) can be rewritten as the
operator identity (acting on arbitrary $t_3$):
\be
[s(t_1),s(t_2)] = s([t_1,t_2]) \; ,
\ee
that we have already shown directly for the grafting operator $s_t$.
Thus the two descriptions are equivalent, as expected.

\subsection{Matrix Representations}
\label{sec:matrix}
Graphs are graded by their number of loops, but there is no canonical
ordering within each graded subspace. Given some
ordering of graphs, the grafting operator $s_t$ can be represented as
a matrix. This matrix will be indexed by graphs.
Now, graphs are inconvenient to use as subscripts, so we use a bra-ket
notation instead: represent $s(t_1)t_2$ by
\[
s(t_1)\ket{t_2} = \sum_{t_3} \ket{t_3} \bra{t_3} s(t_1)\ket{t_2}
\]
where all graphs are to be thought of as orthogonal: 
$\langle {t_1}| {t_2} \rangle =0$
for $t_1 \neq t_2$. 

An example: the insertion tables (consulting appendix \ref{app:E}) gives
\[
s(\pv{\fmfreuse{v1}}) \, \pv{\fmfreuse{v1}} =
{2 \over 3} \, \pv{\fmfreuse{v4}}  + {2 \over 3} \, \pv{\fmfreuse{v5}}  + 
{2 \over 3} \, \pv{\fmfreuse{v6}} \; ,
\]
which is to be compared to the expansion in a complete set
\[
s(\pv{\fmfreuse{v1}}) \, \ket{\ps{\fmfreuse{v1}}} =
\ldots+
\bra{\pv{\fmfreuse{v4}}} s(\pv{\fmfreuse{v1}}) \ket{\pv{\fmfreuse{v1}}}
\;  \ket{\pv{\fmfreuse{v4}}} + 
\bra{\pv{\fmfreuse{v5}}} s(\pv{\fmfreuse{v1}}) \ket{\pv{\fmfreuse{v1}}}
\;  \ket{\pv{\fmfreuse{v5}}} + 
\bra{\pv{\fmfreuse{v6}}} s(\pv{\fmfreuse{v1}}) \ket{\pv{\fmfreuse{v1}}}
\;  \ket{\pv{\fmfreuse{v6}}}  + \ldots
\]
so we read off
\[
\bra{\pv{\fmfreuse{v4}}} s(\pv{\fmfreuse{v1}})
\ket{\pv{\fmfreuse{v1}}}=  2/3 \; , \quad
\bra{\pv{\fmfreuse{v5}}} s(\pv{\fmfreuse{v1}})
\ket{\pv{\fmfreuse{v1}}}= 2/3 \; , \quad
\bra{\pv{\fmfreuse{v6}}} s(\pv{\fmfreuse{v1}})
\ket{\pv{\fmfreuse{v1}}}= 2/3 \; .
\]
giving the second column of the matrix:
\renewcommand{\arraystretch}{0.7}
\setlength{\arraycolsep}{1.2mm}
\be
s(\pv{\fmfreuse{v1}})\pv{\fmfreuse{v1}} \rep 
\left( \ba{cccccc}
0 & 0 & 0 & 0 & 0 & 0 \\
\!1/3\! & 0 & 0 & 0 & 0 & 0  \\
0 & \!2/3\! & 0 & 0 & 0 & 0  \\
0 & \!2/3\! & 0 & 0 & 0 & 0  \\
0 & \!2/3\! & 0 & 0 & 0 & 0  \\
0 & 0 & 0 & 0 & 0 & 0  \\
\ea \right) 
\left( \ba{c} 0 \\ 1 \\ 0 \\ 0 \\ 0 \\ 0 \ea \right)
= \left( \ba{c} 0 \\ 0 \\ \! 2/3 \! \\ \! 2/3 \! \\ \! 2/3 \!
 \\ 0  \ea \right)
\rep
{2 \over 3} \, \pv{\fmfreuse{v4}}  +{2 \over 3}  \, \pv{\fmfreuse{v6}}  + 
{2 \over 3} \, \pv{\fmfreuse{v5}}
\label{eq:sv1}
\ee
\renewcommand{\arraystretch}{1}
where we have ordered the graphs as
\pv{\fmfreuse{v0}},\pv{\fmfreuse{v1}}, \pv{\fmfreuse{v4}}, 
\pv{\fmfreuse{v5}},\pv{\fmfreuse{v6}}, \pv{\fmfreuse{v7}},
and $\rep$ denotes ``represents''. 
In the appendix we give first insertion tables up to three-loop order
(appendix \ref{app:E}),
then a few examples of matrices that represent the grafting
operators (appendix \ref{app:F}). 
These matrices are easily extracted from the insertion
tables, and are indexed by graphs.

A word about practical implementation: because of our space-saving
convention of suppressing graphs related by crossing symmetry, each
number in the above $6\times 6$ matrix is a $3 \times 3$ unit matrix.
The exception is the first column, which has as second index the 
single tree-level diagram and so is a $3 \times 1$ matrix. 
Thus, in a completely
explicit notation, the above matrix is $16 \times 16$.

A few remarks are in order.
It is immediately clear that the matrices $s_t$
are all lower triangular, since insertion can never decrease loop number.
(Recall that $s_1$, which is represented by a diagonal matrix,
is left out of the algebra). Triangularity
makes these matrices 
easy to exponentiate, since the series will cut off at finite order. 
This property will be crucial in section \ref{renormalization}.
Triangular matrices are non-invertible,
which makes sense---typically each application of $s_t$ creates many
different graphs. There is no unique inverse under $s_t$ of a generic
graph (but
see section \ref{sec:shrinking}). 
Finally, by a quick glance 
in the appendix we see that the matrices are very
sparse, which is useful to know for computer storage; 
the size of the matrix is the number of relevant diagrams
squared, which quickly becomes prohibitive if sparsity is not exploited.

It will be useful in the following to
consider exponentiating matrices, for example the matrix 
representing s(\pv{\fmfreuse{v1}}) in (\ref{eq:sv1}):
\bean
\exp(s(\pv{\fmfreuse{v1}}))\pv{\fmfreuse{v0}}
&=& (1 + s(\pv{\fmfreuse{v1}}) +
\frac{1}{2} 
s(\pv{\fmfreuse{v1}}) s(\pv{\fmfreuse{v1}})) \pv{\fmfreuse{v0}}
 \\
&=& \pv{\fmfreuse{v0}} + {1 \over 3}\, \pv{\fmfreuse{v1}}+
{2 \over 3} \, \pv{\fmfreuse{v4}}  + {2 \over 3} \, \pv{\fmfreuse{v6}}  + 
{2 \over 3} \, \pv{\fmfreuse{v5}}
\eean
since the matrix (\ref{eq:sv1}) vanishes when cubed. On the other hand,
\[
\exp(s(\ps{\fmfreuse{s1}}))\ps{\fmfreuse{v0}} = 0 \; .
\]

Thus, the exponential
of the sum of all grafting operators acts as
\[
e^{\sum_t s_t} \; \pv{\fmfreuse{v0}}
=  \pv{\fmfreuse{v0}} +{1 \over 3}\,  \pv{\fmfreuse{v1}} +
{2 \over 3} \, \pv{\fmfreuse{v4}}  + {2 \over 3} \, \pv{\fmfreuse{v6}}  + 
{2 \over 3} \, \pv{\fmfreuse{v5}} + \ldots
\]
where the dots denote higher-order terms.
Acting with the inverse is now a simple matter:
\be
e^{-\sum_t s_t}\; \pv{\fmfreuse{v0}}
=  \pv{\fmfreuse{v0}} -{1 \over 3}\,  \pv{\fmfreuse{v1}} +
{2 \over 3} \, \pv{\fmfreuse{v4}}  + {2 \over 3} \, \pv{\fmfreuse{v6}}  + 
{2 \over 3} \, \pv{\fmfreuse{v5}} + \ldots
\label{eq:inverse}
\ee
With the three-loop matrices given in the appendix,
one easily performs exponentiation to three-loop level as well (see
section \ref{sec:threeloop}).
The alternating signs in equation (\ref{eq:inverse})
is already now suggestive for the reader who is familiar with Hopf
algebra.
The antipode of the Hopf algebra has a similar alternating sign---each
shrinking of a subgraph comes with a sign change.
In section \ref{renormalization} we display
the relation to the antipode, and thus the counterterms of the theory.

\section{Other Operations}
\subsection{Shrinking}
\label{sec:shrinking}
When acting with the transpose of the matrix representation of a
grafting operator, we shrink graphs.
Here is an example:
\renewcommand{\arraystretch}{0.7}
\setlength{\arraycolsep}{1.2mm}
\[
s^T(\pv{\fmfreuse{v1}})\pv{\fmfreuse{v4}} \rep 
\left( \ba{cccccc}
0 & 1/3 & 0 & 0 & 0 & 0 \\
0 & 0 & 2/3 & 2/3 & 2/3 & 0  \\
0 & 0 & 0 & 0 & 0 & 0  \\
0 & 0 & 0 & 0 & 0 & 0  \\
0 & 0 & 0 & 0 & 0 & 0  \\
0 & 0 & 0 & 0 & 0 & 0  \\
\ea \right) 
\left( \ba{c} 0 \\ 0 \\ 1 \\ 0 \\ 0 \\ 0 \ea \right)
= \left( \ba{c} 0 \\ 2/3 \\ 0 \\ 0 \\ 0 \\ 0 \ea \right)
\rep 2/3 \, \pv{\fmfreuse{v1}}
\]
\renewcommand{\arraystretch}{1}
This operation does not have a counterpart in the star notation;
we will not explore it further here.

\subsection{Gluing and External Leg Corrections}
\label{sec:gluing}
By insertion we only create 1-particle-irreducible (1PI)
diagrams. We can also define the
operation of {\it gluing}, that is, joining any two external legs of
two graphs with a new propagator. Under this operation, either the
insertion graph
becomes an external leg correction to the object graph (if the
insertion is has 2 external legs)
or we add more external legs (if the insertion has
$n$ external legs with $n>2$). In the latter case, we obtain
$n$-point functions with $n>4$, 
which are not superficially divergent in $\phi^4$
theory, so we do not consider them. 
As for the external leg correction, it 
does not have to be considered here either, 
since we only need amputated graphs for the $S$-matrix. 

Thus, the gluing operation will not be of much direct
interest to us, but it does have one property we wish to emphasize.
Define $t_1 \circ t_2$ to be gluing of $t_1$ onto $t_2$, e.g.\
$\pv{\fmfreuse{v1}} \circ \ps{\fmfreuse{s1}}$ is an external leg
correction to $\pv{\fmfreuse{v1}}$.
The operator $s$ satisfies the Leibniz rule on these diagrams:
\be
s(t_1)(t_2 \circ t_3) = ( s(t_1) t_2) \circ t_3+
t_2 \circ (s(t_1)t_3) \; .
\label{eq:gluing}
\ee
It is easy to check that this relation holds in examples in $\phi^4$
theory. 

The structure of (\ref{eq:gluing}) also seems to indicate a possible
interpretation as a 
coderivation. We will not explore this direction in this work, since as
we have seen, the gluing operation is not needed for computing
$S$-matrix elements in our model field theory.

\section{Renormalization}
\label{renormalization}
Consider a matrix $M_1$ depending on a parameter $\epsilon$,
with elements that
diverge when $\epsilon \rightarrow 0$. 
Writing
$M_1(1/\epsilon, \epsilon)$, we mean that both positive and negative
powers of $\epsilon$ occur in the matrix elements. If
only $\epsilon$ occurs in the
argument and not $1/\epsilon$, there are no negative powers. 
We can decompose
the divergent matrix $M_1$ into a finite part $M_2$ and a divergent part
$M_3$ by the general procedure known as Birkhoff decomposition:
\be
M_1(1/\epsilon, \epsilon) = M_2(1/\epsilon) M_3(\epsilon) \; ,
\label{eq:birkhoff}
\ee
which is uniquely fixed by a boundary condition $M_2(0)=0$.
The matrix $M_3$ has a limit as $\epsilon \rightarrow 0$.
Birkhoff decomposition was applied to renormalization
by Connes and Kreimer \cite{CK2}. 
We take a somewhat different approach; those
authors treat the three matrices in the decomposition
(\ref{eq:birkhoff}) 
as the same type of objects, 
whereas we will use the Lie algebra to reduce 
$M_1$ and $M_3$ to vectors, but keep $M_2$ as a matrix.
The main point is that we choose $M_2$ in the group
corresponding to the Lie algebra of matrices defined in section
\ref{sec:matrix}. 
That is, the divergent matrix $M_2$ is a matrix $C$ of the form
\[
C = \exp \left( \sum_t C(t) s(t) \right)
\]
with numerical coefficients $C(t)$ representing the overall
divergence of the graphs $t$. 
In dimensional regularization,
$\epsilon$ represents $\epsilon=4-D$ where $D$ is the (complex) 
spacetime dimension. 
The boundary condition $M_2(0)=0$ is realized in the MS scheme, but
other schemes can be accommodated by adapting other boundary
conditions. 
The coefficient $C(t)$ will depend on $\epsilon$ so
it should properly be denoted $C_{\epsilon}(t)$, however, we will
suppress this dependence.
The coefficient $C(t)$
 is a polynomial in $1/ \epsilon$ with no constant term (again, this
is the boundary condition for (\ref{eq:birkhoff})), 
but can depend on external momenta. 
To calculate the complete set of counterterms, it suffices to know these
overall-divergence coefficients. 

Let us describe the renormalization procedure in this framework, 
assuming we know the matrix $C$,
i.e.\ both the overall-divergence
values $C(t)$ and the combinatorial matrices $s_t$.
Denote the bare value of a graph
$t$ by $B(t)$ and the renormalized graph by $A(t)$ (in the
graph-by-graph method \cite{Collins}). 
Vectors containing these values, indexed by
graphs, are denoted $A$ and $B$, respectively.
We Birkhoff-decompose the vector of bare values
as in equation (\ref{eq:birkhoff}):
\[
B_{1/\epsilon, \epsilon} = C_{1/\epsilon}\,  A_{\epsilon} \; .
\]
To find renormalized values from bare values, we 
have to invert the matrix $C$,
which is a trivial matter when it is expressed as an exponential:
\bea
A_{\epsilon} &=& (C_{1/\epsilon})^{-1} B_{1/\epsilon, \epsilon} 
\nonumber \\
&=&  
\exp \left(- \sum_t C(t) s(t) \right) B_{1/\epsilon, \epsilon} \; .
\label{eq:ACB}
\eea
This is our main statement.
As we already pointed out, the sign in the exponential reproduces the
sign of Zimmermann's forest formula \cite{BPHZ}, 
since every $s(t)$ factor in the expansion of
the exponential corresponds to
an insertion, and a sign.

This is most easily understood in an example.
Let us calculate the renormalized 4-point function up to two loops:
\bean
\exp\left(-\sum_t C(t) s(t)\right) B
&=& 
\left(
\ba{cccccc}
 1 & 0 & 0 & 0 & 0 & 0 \\
-C(\pv{\fmfreuse{v1}}) & 1 & 0 & 0 & 0 & 0 \\
-C(\pv{\fmfreuse{v4}})+2 C(\pv{\fmfreuse{v1}})C(\pv{\fmfreuse{v1}})
 & -2C(\pv{\fmfreuse{v1}}) & 1 & 0 & 0 & 0 \\
-C(\pv{\fmfreuse{v5}})+C(\pv{\fmfreuse{v2}}\, )C(\pv{\fmfreuse{v1}})
 & -C(\pv{\fmfreuse{v1}}) & 0 & 1 & 0 & 0 \\
-C(\pv{\fmfreuse{v6}})+ C(\pv{\fmfreuse{v3}}\, )C(\pv{\fmfreuse{v1}} )
 & -C(\pv{\fmfreuse{v1}}) & 0 & 0 & 1& 0  \\
-C(\pv{\fmfreuse{v7}})+ C(\pv{\fmfreuse{v1}})C(\ps{\fmfreuse{s1}})
 & -C(\ps{\fmfreuse{s1}}) & 0 & 0 & 0 & 1 \\
\ea \right)
\left(
\ba{c}
B(\pv{\fmfreuse{v0}}) \\
B(\pv{\fmfreuse{v1}}) \\
B(\pv{\fmfreuse{v4}}) \\
B(\pv{\fmfreuse{v5}}) \\
B(\pv{\fmfreuse{v6}}) \\
B(\pv{\fmfreuse{v7}}) \\
\ea \right)
\eean
where we, as mentioned earlier, 
it is convenient to use a symmetric renormalization point
so that $C(\pv{\fmfreuse{v1}})=C(\pv{\fmfreuse{v2}} \,)
=C(\pv{\fmfreuse{v3}}\, )$. In fact, these three
graphs always appear in the same place in the expansion, since
$s(\pv{\fmfreuse{v1}})=s(\pv{\fmfreuse{v2}}\, )
=s(\pv{\fmfreuse{v3}}\, )$.
We should also emphasize that juxtaposition of $C(t)$ cannot be interpreted as
multiplication. Instead, looking at the matrix $C^{-1}$, a term such
as $C(\pv{\fmfreuse{v2}}\, )C(\pv{\fmfreuse{v1}})$ corresponds to
$(1/2)
\langle\langle \pv{\fmfreuse{v2}}\, \rangle \pv{\fmfreuse{v1}}\rangle $
where the angle brackets
$\langle \; \rangle$ pick out the pole term in minimal subtraction
(in general, the $1/2$ is $1/N!$).

We have, for the nontrivial elements,
\bean
\left( \ba{c}
A(\pv{\fmfreuse{v1}}) \\
A(\pv{\fmfreuse{v4}}) \\
A(\pv{\fmfreuse{v5}}) \\
A(\pv{\fmfreuse{v6}}) \\
A(\pv{\fmfreuse{v7}}) 
\ea  \right)
=
\left(
\ba{c}
-C(\pv{\fmfreuse{v1}}) B(\pv{\fmfreuse{v0}})
+B(\pv{\fmfreuse{v1}}) \\
-(C(\pv{\fmfreuse{v4}})-2 C(\pv{\fmfreuse{v1}})C(\pv{\fmfreuse{v1}}))
B(\pv{\fmfreuse{v0}})
-2C(\pv{\fmfreuse{v1}})
B(\pv{\fmfreuse{v1}})+ B(\pv{\fmfreuse{v4}}) \\
-(C(\pv{\fmfreuse{v5}})- C(\pv{\fmfreuse{v2}}\, )C(\pv{\fmfreuse{v1}} )) 
B(\pv{\fmfreuse{v0}})
-C(\pv{\fmfreuse{v1}})
B(\pv{\fmfreuse{v1}})+B(\pv{\fmfreuse{v5}}) \\
-(C(\pv{\fmfreuse{v6}})- C(\pv{\fmfreuse{v3}}\, )C(\pv{\fmfreuse{v1}} ))
 B(\pv{\fmfreuse{v0}})
-C(\pv{\fmfreuse{v1}})
B(\pv{\fmfreuse{v1}})+B(\pv{\fmfreuse{v6}}) \\
-(C(\pv{\fmfreuse{v7}})-C(\pv{\fmfreuse{v1}})C(\ps{\fmfreuse{s1}})) 
B(\pv{\fmfreuse{v0}})
-C(\ps{\fmfreuse{s1}})B(\pv{\fmfreuse{v1}})+B(\pv{\fmfreuse{v7}}) \\
\ea \right)\eean
To be specific, consider the third row:
\be
A(\pv{\fmfreuse{v5}})=
-(C(\pv{\fmfreuse{v5}})- C(\pv{\fmfreuse{v2}}\, )C(\pv{\fmfreuse{v1}})) 
 B(\pv{\fmfreuse{v0}})
- C(\pv{\fmfreuse{v1}}) B (\pv{\fmfreuse{v1}})
+ B(\pv{\fmfreuse{v5}}) \; .
\label{eq:specific}
\ee
The combination of graphs is clearly correct; 
the second counterterm (third term)
cancels the potentially nonlocal divergence in
$\pv{\fmfreuse{v5}}$, and the first counterterm (first and second
terms) takes care of the
overall divergence after the nonlocal one is canceled.
There is an even number of terms, as expected from the underlying
Hopf algebra structure. 
The matrix $C$ ``knew'' the connection between a graph and its
divergent subgraphs, since $C$ was constructed from the 
grafting operators $s_t$. 
We can check the cancellation of nonlocal divergences
in (\ref{eq:specific}). Evaluating now in 
$\phi^4$ theory, with $f_1$, $f_2$, $g_1$, $g_2$ containing no
negative powers of $\epsilon$ and at most polynomial in $p$: 
\bean
A(\pv{\fmfreuse{v5}}) &=&{\lambda^2 \over 2(4\pi)^2}\Bigg\{
-\left[\left({2 \over \epsilon^2}-{2 \over \epsilon}\ln p^2 + {1 \over
\epsilon}\, f_1  \right) 
-\left({2 \over \epsilon^2}-{2 \over \epsilon}\ln p^2
+{2 \over \epsilon}\, g_1 \right) \right] \\
&& - \left(
{2 \over \epsilon^2}-{2 \over \epsilon}\ln p^2 + {2 \over
\epsilon}\, g_1 + 2 g_2 \right)
+\left({2 \over \epsilon^2}-{2 \over \epsilon}\ln p^2 + {1 \over
\epsilon}\, f_1 +f_2 \right) \Bigg\} \\
&=& {\lambda^2 \over 2(4\pi)^2} (f_2-2g_2) \; .
\eean
Here we
included the symmetry factors (see Appendix \ref{app:sym}) 
 of $1/2$.
We see that the cancellation
of $(1/\epsilon)\ln p^2$ is assured.
The second check is that the second order vertex counterterm 
$(C(\pv{\fmfreuse{v5}})- C(\pv{\fmfreuse{v2}}\, )C(\pv{\fmfreuse{v1}}))
\; \propto \; (2g_1-f_2)/\epsilon$ is local.

The total contribution to the 4-point function up to two-loop order is
just the sum of all compatible (4-external-leg) elements of $A$, as in
\bean {\mathcal A} &=& A(\pv{\fmfreuse{v0}})+A(\pv{\fmfreuse{v1}})+
A(\pv{\fmfreuse{v4}})+A(\pv{\fmfreuse{v5}})+A(\pv{\fmfreuse{v6}})
\quad (+\mbox{crossing}) \eean The total combination of graphs can be
read off from the $A$ vector above.

\section{Three-loop Example}
\label{sec:threeloop}
All the diagrams at 3-loop order in $\phi^4$ theory
have been computed, and clever
techniques have been invented to perform the computations.
Some techniques are reviewed in \cite{CT}.
A useful reference for tables of
divergent parts up to three loops is \cite{CC}.
Therefore, we need only worry about the combinatorics,
but for completeness we try to convey 
an idea which integral computation techniques can
be used to compute the relevant diagrams.

In one-loop computations, one usually employs 
the method of Feynman parameters.
In fact, using Feynman parameters without additional techniques
is, by some estimates,
only economical for graphs with two lines. 
``Additional techniques'' can include performing some series expansions
before computing the integral (the Gegenbauer polynomials have been
shown to be well suited for this), and imaginative use of
integration by parts. In this context, integration by parts 
is applied to simplify integrals using the vanishing of integrals over
total derivatives:
\[
\int d^4 p \; {\partial \over \partial p} I(p) = 0 
\] 
for any integrand $I(p)$ depending on the loop momentum $p$ and any other
loop or external momenta. 
Applying integration by parts, 
the massive ladder diagram \pu{\fmfreuse{u2}} 
at zero momentum transfer can be decomposed as \cite{CMM}  \\
\bean
 \pu{\fmfreuse{u2}} &\; \eva \; &{1 \over (4\pi)^4}
\left({m^2 \over 4 \pi M^2}\right)^{ \!\! -\epsilon}
\int {d^D k \over (2\pi)^D} {[F(k^2)]^2 \over (k^2 + m^2)} \\
&&+ 
{1 \over \epsilon} {4 \over (4\pi)^2}
\left({m^2 \over 4 \pi M^2}\right)^{ \!\! -\epsilon/2}  \! \! W_6
- {1 \over \epsilon^2} {4 \over (4\pi)^2}
\left({m^2 \over 4 \pi M^2}\right)^{ \!\! -\epsilon} \!  \!S_3 \; ,
\eean
where $F$ is a simpler genuine three-loop integral,
\ie\ it cannot be expressed in terms of integrals of lower loop order,
$W_6$ is 
a two-loop integral ($\pv{\fmfreuse{v5}}$ at zero momentum transfer)
and $S_3$ is a one-loop integral ($\pv{\fmfreuse{v1}}$
at zero momentum transfer). The result (given in \cite{CC}) is
\bean
 \pu{\fmfreuse{u2}} &\; \eva \; &{m^4 \over (4\pi)^6 }
\left({m^2 \over 4 \pi M^2}\right)^{-3\epsilon /2}
\left(
{8 \over 3 \epsilon^3} +
{1 \over \epsilon^2}
\left[ {8 \over 3} - 4 \gamma \right] \right.
\\[1mm]
&& \qquad \left.
+ {1 \over \epsilon}
\left[ {4 \over 3} - 4a -4 \gamma + 3 \gamma^2 + {\pi^2 \over 6}
 \right]  \right)
\mbox{ + (finite)}
\eean
where $\gamma$ is Euler's constant, 
$a$ is a numerical constant ($a=1.17...$) coming from
integration over Feynman parameters,
and $M$ is the renormalization scale.

Now let us use the matrix Lie algebra to compute the renormalized
graph (again working in the graph-by-graph context). 
From the previous section, we know that counterterms are
generated by the exponential of a sum over grafting operators. 

Using
\bean
A &=& 
\exp\bigg[ -(C(\pv{\fmfreuse{v1}})s(\pv{\fmfreuse{v1}})
+ C(\pv{\fmfreuse{v4}})s(\pv{\fmfreuse{v4}})
+ C(\pv{\fmfreuse{v5}})s(\pv{\fmfreuse{v5}}) + \ldots) \bigg]\,  B \; ,
\eean
we find the renormalized graphs.
In particular, the relevant row of the vector $A$ is
\bea
A(\pu{\fmfreuse{u2}})  
&=& \left( -\fr{2}{81} C(\pv{\fmfreuse{v1}})^3 
+ \fr{1}{18} C(\pv{\fmfreuse{v1}}) C(\pv{\fmfreuse{v5}})
+\fr{1}{18} 
C(\pv{\fmfreuse{v1}}) C(\pv{\fmfreuse{v6}})-\fr{1}{6}
C(\pu{\fmfreuse{u2}}) \right)
B(\pv{\fmfreuse{v0}}) +  \nonumber \\
&&
  \, \fr{2}{9} C(\pv{\fmfreuse{v1}})^2 B(\pv{\fmfreuse{v1}})
- \fr{1}{3} C(\pv{\fmfreuse{v1}}) B(\pv{\fmfreuse{v5}})
-  \fr{1}{3} C(\pv{\fmfreuse{v1}}) B(\pv{\fmfreuse{v6}})
+ B(\pu{\fmfreuse{u2}}) \; .
\label{eq:result}
\eea
This is to be compared to the known result for the renormalized graph:
\bea
 \pu{\fmfreuse{u2}} \bigg|_{\rm ren}  &=&
\pu{\fmfreuse{u2}} + 
\bigg(
-\langle \pu{\fmfreuse{u2}} \rangle + 
\langle \langle \pv{\fmfreuse{v1}} \rangle \pv{\fmfreuse{v5}} \rangle +
\langle \langle \pv{\fmfreuse{v1}} \rangle \pv{\fmfreuse{v6}} \rangle 
-\langle \langle \pv{\fmfreuse{v1}} \rangle^2
 \pv{\fmfreuse{v2}} \,   \rangle \bigg)
\nonumber \\[1mm]
&& 
- \langle \pv{\fmfreuse{v2}} \,  \rangle \pv{\fmfreuse{v5}} 
- \langle \pv{\fmfreuse{v2}} \,  \rangle \pv{\fmfreuse{v6}}  +
  \langle \pv{\fmfreuse{v2}} \,  \rangle ^2   \pv{\fmfreuse{v2}}
\label{eq:known} 
\eea
where $\langle \;  \rangle$ denotes the renormalization map (for
example, in minimal subtraction we simply drop the finite
part). Rewriting the known expression (\ref{eq:known}) 
in language more similar to the above, we have
\bean
A(\pu{\fmfreuse{u2}})&=& B(\pu{\fmfreuse{u2}})
+\left(-C(\pu{\fmfreuse{u2}})+ C(\pv{\fmfreuse{v1}})
C(\pv{\fmfreuse{v5}}) + C(\pv{\fmfreuse{v1}})C(\pv{\fmfreuse{v6}})
-C(\pv{\fmfreuse{v1}})^2 C(\pv{\fmfreuse{v2}}\, ) \right)
B(\pv{\fmfreuse{v0}}) \\[1mm]
&& - C(\pv{\fmfreuse{v2}}\, )B(\pv{\fmfreuse{v5}}) -
C(\pv{\fmfreuse{v2}}\, )B(\pv{\fmfreuse{v6}})
+C(\pv{\fmfreuse{v2}}\, )^2B(\pv{\fmfreuse{v2}}\, ) \; .
\eean
Expression (\ref{eq:known}) is reproduced by
that derived by the Lie algebra, up to the 
differing normalization (see section \ref{sec:grafting}).
To summarize,
we have used the inverse exponential of Lie algebra elements to 
generate counterterms, just as the antipode would have been used in a
Hopf algebra.

The Ihara bracket relation $s_{[t_1,t_2]}=[s_{t_1},s_{t_2}]$ is still
rather trivial in this example, as it turns out, because only
$s(\pv{\fmfreuse{v1}})$ appears more than once. 
In the example of \pu{\fmfreuse{u8}}, we have
$s(\pv{\fmfreuse{v1}})$, $s(\pv{\fmfreuse{v4}})$
and $s(\pv{\fmfreuse{v5}})$ appearing, so there is
potential for use of the relation here. 
However, the only nontrivial commutator, equation
(\ref{eq:commut}), yields graphs that do not appear 
as subgraphs of \pu{\fmfreuse{u8}}:
\[
 {1 \over 3} \, s(\ps{\fmfreuse{s2}})+ 
{2 \over 3}\, s(\pv{\fmfreuse{s3}})  
- 2\, s(\pv{\fmfreuse{v7}}) 
= [ s(\pv{\fmfreuse{v1}}), s(\ps{\fmfreuse{s1}})] \; .
\]
We have thus seen that it is not until 
the four-loop level that nontrivial application of 
the Lie algebra relation $s_{[t_1,t_2]}=[s_{t_1},s_{t_2}]$ appear.
As mentioned in the introduction,
new results are needed at five loops and higher.

\section{Renormalization Group Flows
and Nonrenormalizable Theories}
\label{sec:Wilson}
In Wilson's approach to renormalization, 
short-distance (ultraviolet) fluctuations are integrated out
of the functional integral to appear only as modifications of
parameters of the Lagrangian for the remaining long-distance
(infra\-red) degrees of freedom \cite{KW}. 
The introduction of a mass scale $M$ by
renormalization parameterizes different renormalization schemes, 
and the change of $M$ induces a flow of parameters in the Lagrangian,
such that certain so-called irrelevant operators die away.
In $\phi^4$ theory in four dimensions, 
$\phi^6$ is an example of one such operator;
the relative size of this operator to other terms in the Lagrangian
at a momentum scale $p$
would be $(p/\Lambda)^{6-D}=(p/\Lambda)^2$ as $p\rightarrow 0$, where
$\Lambda$ is a cutoff.
 
Now, the grafting operator $s_t$ may be thought of as a ``magnifying
glass''; by inserting graphs, we resolve details of graphs that were not
visible at larger scales. In this specific sense, $s_t$ induces scale
change. 
In particular, in this paper, we only considered $s_t$ for graphs $t$
with 2 or 4 external legs.
 We do not have to consider $s_t$ for graphs with 6 external
legs in $\phi^4$ theory in four dimensions, 
precisely because $\phi^6$ is an irrelevant
operator in this theory. 
This shows how the matrix Lie algebra would appear in a
nonrenormalizable theory; everything is the same as in the $\phi^4$
example,  except there is an infinite number of insertion matrices. 
While this situation is certainly 
more cumbersome, it is not necessarily fatal.

According to Connes and Kreimer \cite{CK2}, renormalization group flow
boils down to the calculation of 
a matrix $\beta$ which in our notation becomes
\[
\beta = \sum_t \beta(t)\, s_t \; ,
\]
and is independent of $\epsilon$.
Here $\beta(t)$ is the numerical
contribution to the beta function due to a certain
diagram $t$ (again, in the graph-by-graph method).
This matrix $\beta$ 
generalizes the standard beta function of the renormalization
group; there is now combinatorial information attached to
each of the contributions $\beta(t)$ to the beta function. 

See also the next section for some comments on the 
range of validity
of Wilson's point of view.

\section{Conclusion and Outlook}
\label{sec:div}
In this work, we displayed a matrix Lie algebra of 
operators acting on Feynman graphs, that by
exponentiation yields group elements generating
counterterms for diagrams with subdivergences. 
We defined a grafting operator $s_t$ that inserts one
Feynman graph $t$
into another. The matrix representations of these operators satisfy a
certain rule, similar to an Ihara bracket,
that gives relations between them: $s_{[t_1,
t_2]}=[s_{t_1}, s_{t_2}]$. In this way, not all matrices have to be
separately computed, with potentially substantial savings in computation time.
We displayed the relation to the star product of Kreimer, which is
defined similarly to (but not exactly the same way as) the grafting
operator. A simple computation verifies that
the right-symmetric (pre-Lie) nonassociative
algebra of the star product is equivalent to the
previously mentioned Ihara-bracket rule for our matrix representations. 

We also gave a number of examples, mostly rather simple ones, and
checked in a three-loop example
that the correct sequence of counterterms is provided by the
exponential of Lie algebra elements.  
(The general proof that this is always the case will be given
elsewhere \cite{Cartier}). 
Just as with Hopf algebra, the Lie algebra rules are trivial at
one-loop, almost trivial at two-loop, and just beginning to become
interesting at three-loop order. At four loops, there is plenty of
interesting structure to check, but it will require 
a fair amount of computation; 
this is an obvious direction in which future work
should go.

An equally obvious, but less direct, application is to noncommutative
geometry. Using the Connes-Kreimer homomorphism between the
Hopf algebra of renormalization
and the Hopf algebra of coordinates on the group of formal
diffeomorphisms \cite{CK2,CM}, and the Milnor-Moore theorem,
the Lie algebra in this paper has a
corresponding Lie algebra in the context of noncommutative
geometry. We hope that the matrix 
Lie algebra may eventually shed some
light on certain aspects of noncommutative geometry, in this rather indirect
but novel way.

In a less obvious direction,
it is suggestive to note that the grafting operator
$s_t$ is an element of a Lie algebra, it satisfies a 
Leibniz rule (the gluing
operator, section \ref{sec:gluing}),
and its bracket (\ref{eq:ihara}) looks like that of a
vector field $X$ on a manifold. 
Loosely, the 
grafting operator is a ``scale vector'', that takes us
into ever increasing magnification of graphs by including more and
more subgraphs. If a Lie derivative $\sL_{s_t}$ along this ``vector
field'' could 
meaningfully be defined, that would open the following interesting
speculative possibility. 

It was proposed in \cite{CBDW}, based on earlier
work \cite{CD} that volume forms for functional
integration be characterized the following way: find a vector field
$X$ on the space, define its divergence (this may be a nontrivial
task in an infinite-dimensional space) and let the volume form $\omega$
be defined by 
\[
{\mathcal L}_X \, \omega = ({\rm div}\, X ) \, \omega \; .
\]
This definition reproduces familiar volume forms in some simple
finite-dimen\-sional examples (Riemannian manifolds, symplectic
manifolds), and gives some hope for generalization to
infinite-dimensional spaces.
Perhaps if $s_t$ is a vector field, it
could be used in the role of $X$, in some extended sense, to
characterize $\omega$. In effect, this would be a perturbative definition,
since
including $s_t$ up to a certain loop order will define different
$\omega$ at different orders in the loop expansion. 

An obvious problem with this idea is that
$s_t$ acts on the space of {\it graphs},
not the space of fields. This means that even if
a volume form $\omega$ could be defined, it would be
a function on the space of graphs.
In principle, it may be possible to exploit some analogy to
the space of paths in
quantum-mechanical functional integrals.
This direction would be interesting to pursue in future work.

As a final note, there has been some dispute to what extent Wilson's
picture is generally valid; there are some (as yet speculative)
examples where infrared and ultraviolet divergences are connected
\cite{MRS}. Some of these examples are in connection with
noncommutative geometry. In view of the relation between
the algebraic approach to renormalization and
noncommutative geometry mentioned above, one could hope that
these algebraic developments may eventually shed some light on the
ultraviolet-infrared connection.

\section{Acknowledgments}

We wish to thank C. DeWitt-Morette, 
D. Kreimer and A. Wurm for useful and pleasant discussions.
We would especially like to thank D. Kreimer for feedback on an
early draft of this paper.
MB wishes to thank the Swedish Foundation for
International Cooperation in Research and Higher Education for
financial support, and the 
Departement de Math{\'e}matiques et
Applications, Ecole Normale Sup{\'e}rieure, Paris, for
hospitality.

\newpage

\begin{appendix}

\section{Symmetry factors}
\label{app:sym}

Since conventions for how graphs carry symmetry factors vary,
let us be precise.

The bare graph in our two-loop example contains in its bare value
the nonlocal divergence $(1/\epsilon) \ln p^2$, including
the 1/2 due to the symmetry factor 2.
The counterterms must subtract the $(1/\epsilon) \ln p^2$, 
nothing more, nothing less.

How the notation carries symmetry factors, however, 
varies in the literature. The term 
$\langle \pv{\fmfreuse{v2}} \; \rangle \pv{\fmfreuse{v1}}$ 
taken literally looks like it would contain two symmetry
factors of 2, i.e. the forefactor would be $1/2 \times 1/2 = 1/4$.  
By itself, this is of course wrong ---
it only subtracts half the nonlocal divergence.
Two possible resolutions are
{\it i)} to say that the graph with a cross (as in \cite{Collins}) 
that represents $\langle \pv{\fmfreuse{v2}}\;  \rangle \pv{\fmfreuse{v1}}$ 
has only symmetry factor 2, not 4 (which is not quite consistent notation
on the non-graph level without a factor of 2, if 
$\langle \; t \; \rangle $  is to really mean the
pole part of $t$), or
{\it ii)} to posit, as we do here,
that insertion can also create a term with $C(\pv{\fmfreuse{v3}})$, 
that combines with
$C(\pv{\fmfreuse{v2}}\; )$ when we take
the counterterms to be the same, to give a factor of 2.

It is a useful check to consider the $s$-channel
counterterm as well. In contrast to the previous example,
the graph with an $s$-channel insertion (the first $L=2, E=4$ graph in our
table 1) has a forefactor of 1/4. Each counterterm also has 1/4, but now
there are two of them, one for shrinking each of the two disjoint
subdivergences. Thus they do not cancel directly, but factor as usual.
In the Hopf-algebra 
coproduct, factorization produces the factor of 2.
The antipode then yields a factor of 2
similar to the above. With these conventions, finding a factor of 2 both
in the $s$ and $t/u$ cases, one would expect the $s/t/u$ graphs to
have the same normalization, after summing.

\section{Insertion Tables}
\label{app:E}
In this appendix, we provide tables of the action of the grafting
operator $s_t$.
We have defined $s_t$ such that on a bare diagram denoted by $1$,
we have $s_t 1 = t$. The matrix $s_1$ acts as 
$s_1 t = N(t) t$ where $N(t)$ is the number of insertion points 
of $t$. It is also true that the coefficients of
each resulting graph in an insertion into a graph $t$ should add up to
the number of compatible insertion points of the graph $t$.
\bean
\mbox{One-loop:} \\
1+ 0 = 1\\[2mm]
s(\ps{\fmfreuse{s1}}) \, \pv{\fmfreuse{s0}} &=&  \, \ps{\fmfreuse{s1}} \\
s(\ps{\fmfreuse{s1}}) \, \pv{\fmfreuse{v0}} &=& 0 \\
s(\pv{\fmfreuse{v1}}) \, \pv{\fmfreuse{s0}} &=& 0 \\ 
s(\pv{\fmfreuse{v1}}) \, \pv{\fmfreuse{v0}} &=& 
{1 \over 3} \, \pv{\fmfreuse{v1}} \; +  \mbox{ crossing} \\ [2mm]
\mbox{Two-loop:} \\
1 + 1 = 2\\[2mm]
s(\ps{\fmfreuse{s1}}) \, \ps{\fmfreuse{s1}} &=&  \, \ps{\fmfreuse{s2}} \\
s(\ps{\fmfreuse{s1}}) \, \pv{\fmfreuse{v1}} &=&  \;2 \, \pv{\fmfreuse{v7}} \\
s(\pv{\fmfreuse{v1}}) \, \ps{\fmfreuse{s1}} &=&
{1 \over 3} \, \ps{\fmfreuse{s2}} + {2 \over 3} \, \pv{\fmfreuse{s3}} \\
s(\pv{\fmfreuse{v1}}) \, \pv{\fmfreuse{v1}} &=&
{2 \over 3} \, \pv{\fmfreuse{v4}}  + {2 \over 3} \, \pv{\fmfreuse{v5}}  + 
{2 \over 3} \, \pv{\fmfreuse{v6}} \\ [2mm]
\mbox{Three-loop:} \\
1 + 2 = 3\\[2mm]
s(\ps{\fmfreuse{s1}}) \, \ps{\fmfreuse{s2}} &=&  
2 \,  \ps{\fmfreuse{s4}} +  \, \ps{\fmfreuse{s5}} \\  [2mm]
s(\ps{\fmfreuse{s1}}) \, \pv{\fmfreuse{v4}} &=& 
 4\,  \pu{\fmfreuse{u18}} \\  [2mm]
s(\ps{\fmfreuse{s1}}) \, \pv{\fmfreuse{v5}} &=& 
2 \, \pu{\fmfreuse{u19}} + 2 \,   \pu{\fmfreuse{u21}}  \\ [2mm]
s(\ps{\fmfreuse{s1}}) \, \pv{\fmfreuse{v6}} &=&
2 \, \pu{\fmfreuse{u20}} + 2  \,  \pu{\fmfreuse{u22}}  \\ [2mm]
s(\ps{\fmfreuse{s1}}) \, \pv{\fmfreuse{v7}} &=&
{4 \over 3} \, \pu{\fmfreuse{u15}} + {4 \over 3}  \,  \pu{\fmfreuse{u16}}
+ {4 \over 3} \, \pu{\fmfreuse{u17}} \\ [2mm]
s(\ps{\fmfreuse{v1}}) \, \ps{\fmfreuse{s2}} &=&  
{2\over 3} \,  \ps{\fmfreuse{s5}} +  {2\over 3} \, \ps{\fmfreuse{s7}}
+ {2\over 3} \, \ps{\fmfreuse{s8}} \\  [2mm]
s(\ps{\fmfreuse{v1}}) \, \ps{\fmfreuse{s3}} &=&  
2 \,  \ps{\fmfreuse{s6}}   \\  [2mm]
s(\pv{\fmfreuse{v1}}) \, \pv{\fmfreuse{v4}} &=& 
{2 \over 3} \, \pu{\fmfreuse{u4}} +   \pu{\fmfreuse{u5}} 
+ {2 \over 3} \, \pu{\fmfreuse{u8}} + {2 \over 3}
\pu{\fmfreuse{u9}} \\  [2mm]
s(\pv{\fmfreuse{v1}}) \, \pv{\fmfreuse{v5}} &=& 
{1 \over 3} \, \pu{\fmfreuse{u3}} + {1 \over 3}  \pu{\fmfreuse{u4}} 
+ {2 \over 3} \, \pu{\fmfreuse{u6}} + {1 \over 3}  \pu{\fmfreuse{u8}}
+ {2 \over 3} \, \pu{\fmfreuse{u10}} + {2 \over 3}\, 
\pu{\fmfreuse{u12}}  \\ [2mm]
s(\pv{\fmfreuse{v1}}) \, \pv{\fmfreuse{v6}} &=&
{1 \over 3}\, \pu{\fmfreuse{u3}} + {1 \over 3}  \pu{\fmfreuse{u4}} 
+ {2 \over 3} \, \pu{\fmfreuse{u7}} + {1 \over 3} \pu{\fmfreuse{u9}}
+ {2 \over 3} \, \pu{\fmfreuse{u11}} + {2 \over 3}
  \pu{\fmfreuse{u13}}  \\ [2mm]
s(\pv{\fmfreuse{v1}}) \, \pv{\fmfreuse{v7}} &=&
{2 \over 3} \, \pu{\fmfreuse{u14}} + {1 \over 3} \, \pu{\fmfreuse{u15}} 
+ {2 \over 3} \, \pu{\fmfreuse{u18}} + {2 \over 3} \, \pu{\fmfreuse{u19}}
+ {2 \over 3} \, \pu{\fmfreuse{u20}} \\ [2mm]
2 + 1 = 3 \\ [2mm]
s(\ps{\fmfreuse{s2}}) \, \ps{\fmfreuse{s1}} &=&  
 \,  \ps{\fmfreuse{s5}} \\  [2mm]
s(\ps{\fmfreuse{s3}}) \, \ps{\fmfreuse{s1}} &=&  
 \,  \ps{\fmfreuse{s7}} \\  [2mm]
s(\ps{\fmfreuse{s2}}) \, \pv{\fmfreuse{v1}} &=&
2 \, \pu{\fmfreuse{u15}}  \\ [2mm]
s(\ps{\fmfreuse{s3}}) \, \pv{\fmfreuse{v1}} &=&
2  \, \pu{\fmfreuse{u14}} \\ [2mm]
s(\pv{\fmfreuse{v4}}) \, \ps{\fmfreuse{s1}} &=&  
{1 \over 3} \,  \ps{\fmfreuse{s5}}+ 
{2 \over 3} \,  \ps{\fmfreuse{s6}} \\  [2mm]
s(\pv{\fmfreuse{v5}}) \, \ps{\fmfreuse{s1}} &=&  
{2 \over 3} \,  \ps{\fmfreuse{s6}}+ {1 \over 6}\,  \ps{\fmfreuse{s7}}+  
{1 \over 6} \,  \ps{\fmfreuse{s8}}\\  [2mm]
s(\pv{\fmfreuse{v6}}) \, \ps{\fmfreuse{s1}} &=&  
{2 \over 3} \,  \ps{\fmfreuse{s6}}+ {1 \over 6}\,  \ps{\fmfreuse{s7}}+  
{1 \over 6} \,  \ps{\fmfreuse{s8}}\\  [2mm]
s(\pv{\fmfreuse{v7}}) \, \ps{\fmfreuse{s1}} &=&  
{1 \over 3} \,  \ps{\fmfreuse{s4}} + {2 \over 3}
\,  \ps{\fmfreuse{s8}} \\  [2mm]
s(\pv{\fmfreuse{v4}}) \, \pv{\fmfreuse{v1}} &=&
{2 \over 3} \, \pu{\fmfreuse{u5}} + {2 \over 3} \, \pu{\fmfreuse{u6}} 
+ {2 \over 3} \, \pu{\fmfreuse{u7}} \\ [2mm]
s(\pv{\fmfreuse{v5}}) \, \pv{\fmfreuse{v1}} &=&
{1 \over 3} \, \pu{\fmfreuse{u4}} + {1 \over 6} \, \pu{\fmfreuse{u8}} 
+ {1 \over 6} \, \pu{\fmfreuse{u9}} 
+ {1 \over 3} \, \pu{\fmfreuse{u10}}  + {1 \over 3} \, \pu{\fmfreuse{u11}} 
+ {1 \over 3} \, \pu{\fmfreuse{u12}}  + {1 \over 3} \, 
\pu{\fmfreuse{u12}}  \\ [2mm]
s(\pv{\fmfreuse{v6}}) \, \pv{\fmfreuse{v1}} &=&
{1 \over 3} \, \pu{\fmfreuse{u4}} + {1 \over 6} \, \pu{\fmfreuse{u8}} 
+ {1 \over 6} \, \pu{\fmfreuse{u9}} 
+ {1 \over 3} \, \pu{\fmfreuse{u10}}  + {1 \over 3} \, \pu{\fmfreuse{u11}} 
+ {1 \over 3} \, \pu{\fmfreuse{u12}}  + {1 \over 3} \, \pu{\fmfreuse{u12}} 
\eean

\section{Grafting Matrices}
\label{app:F}
In this appendix, we list matrix representations
of the grafting operators $s(t)$. 
Since insertion into an $n$-point function can only create an
$n$-point function, we consider submatrices of each separately, and
denote them by subscripts. 
For example, if we call
$s(\ps{\fmfreuse{v1}}){\ps{\vspace{4mm}\fmfreuse{s0}}}=A$ 
and
$s(\pv{\fmfreuse{v1}}){\pv{\vspace{4mm}\fmfreuse{v0}}} =B$,
then the complete matrix representing $s(\pv{\fmfreuse{v1}})$
is the direct sum of $A$ and $B$:
\[
s(\pv{\fmfreuse{v1}}) \rep 
\left( 
\ba{c|c}
A & 0 \\ \hline
0 & B \\ 
\ea \right) \; .
\]
Lines are drawn through
the matrices to delineate loop order $L$.

\ \\ \ \\
\renewcommand{\arraystretch}{0.9}
One-loop insertions:
\setlength{\arraycolsep}{1.4mm}
\[
s(\ps{\fmfreuse{s1}}){\ps{\vspace{4mm}\fmfreuse{s0}}} =
\left( \ba{c|c|cc}
0 & 0 & 0 & 0 \\ \hline
1 & 0 & 0 & 0 \\ \hline
0 & 1 & 0 & 0 \\ 
0 & 0 & 0 & 0 \\
\ea \right)
\ba{c} 
\ps{\fmfreuse{s0}}  \\ \ps{\fmfreuse{s1}}  \\
\ps{\fmfreuse{s2}}  \\ \ps{\fmfreuse{s3}}  \\
\ea \qquad
s(\pv{\fmfreuse{v1}}){\ps{\vspace{4mm}\fmfreuse{s0}}} =
\left( \ba{c|c|cc}
0 & 0 & 0 & 0 \\ \hline
0 & 0 & 0 & 0 \\ \hline
0 & \frac{1}{3} & 0 & 0 \\
0 & \frac{2}{3} & 0 & 0 \\
\ea \right)
\ba{c} 
\ps{\fmfreuse{s0}}  \\ \ps{\fmfreuse{s1}}  \\
\ps{\fmfreuse{s2}}  \\ \ps{\fmfreuse{s3}}  \\
\ea
\]
\[
s(\ps{\fmfreuse{s1}}){\pv{\vspace{4mm}\fmfreuse{v0}}} =
\left( \ba{c|c|cccc} 
0 & 0 & 0 & 0 & 0 & 0  \\ \hline
0 & 0 & 0 & 0 & 0 & 0  \\ \hline
0 & 0 & 0 & 0 & 0 & 0  \\
0 & 0 & 0 & 0 & 0 & 0  \\
0 & 0 & 0 & 0 & 0 & 0  \\
0 & 2 & 0 & 0 & 0 & 0  \\
\ea \right)
\ba{c} 
\pv{\fmfreuse{v0}}  \\ \pv{\fmfreuse{v1}}  \\
\pv{\fmfreuse{v4}}  \\ \pv{\fmfreuse{v5}}  \\
\pv{\fmfreuse{v6}}  \\ \pv{\fmfreuse{v7}}  \\
\ea \qquad
s(\pv{\fmfreuse{v1}}){\pv{\vspace{4mm}\fmfreuse{v0}}} =
\left( \ba{c|c|cccc}
0 & 0 & 0 & 0 & 0 & 0  \\ \hline
\frac{1}{3} & 0 & 0 & 0 & 0 & 0  \\ \hline
0 & \frac{2}{3} & 0 & 0 & 0 & 0  \\
0 & \frac{2}{3} & 0 & 0 & 0 & 0  \\
0 & \frac{2}{3} & 0 & 0 & 0 & 0  \\
0 & 0 & 0 & 0 & 0 & 0  \\
\ea \right)
\ba{c} 
\pv{\fmfreuse{v0}}  \\ \pv{\fmfreuse{v1}}  \\
\pv{\fmfreuse{v4}}  \\ \pv{\fmfreuse{v5}}  \\
\pv{\fmfreuse{v6}}  \\ \pv{\fmfreuse{v7}}  \\
\ea\]
\ \\

When it comes to three-loop matrices, it is typographically easier to
give the transpose:
\setlength{\arraycolsep}{1.2mm}
\bean
s^{\rm T}(\pv{\fmfreuse{v1}}){\pv{\vspace{4mm}\fmfreuse{v0}}} &=&
\left(
 \ba{cccccccccccccccccccccc}
 0 & 0 & 0 & 0 & 0 & 0 & 0 & 0 & 0 &
0 & 0 & 0 & 0 & 0 & 0 & 0 & 0 & 0 & 0 & 0 & 0 & 0  \\ \hline
 0 & 0 & 0 & 0 & 0 & 0 & 0 & 0 & 0 &
0 & 0 & 0 & 0 & 0 & 0 & 0 & 0 & 0 & 0 & 0 & 0 & 0  \\ \hline
 0 & 0 & 0 & \fr{2}{3} & 1 & 0
& 0 & \fr{2}{3} & \fr{2}{3} &
0 & 0 & 0 & 0 & 0 & 0 & 0 & 0 & 0 & 0 & 0 & 0 & 0  \\
 0 & 0 & \fr{1}{3} &\fr{1}{3}  & 0
& \fr{2}{3} & 0 & \fr{1}{3} & 0 &
\fr{2}{3} & 0 & \fr{2}{3} & 0 & 0 & 0 & 0 & 0 & 0 & 0 & 0 & 0 & 0  \\
 0 & 0 &  \fr{1}{3} &\fr{1}{3} 
& 0 & 0 & \fr{2}{3} & 0
& \fr{1}{3} & 0 & \fr{2}{3} & 0 & \fr{2}{3} &
0 & 0 & 0 & 0 & 0 & 0 & 0 & 0 & 0  \\
 0 & 0 & 0 & 0 & 0 & 0 & 0 & 0 & 0 &
0 & 0 & 0 & 0 & \fr{2}{3} & \fr{1}{3} & 0 & 0 & \fr{2}{3} &
\fr{2}{3} & \fr{2}{3} & 0 & 0
\ea  \right) 
\ba{c} 
\pv{\fmfreuse{v0}}  \\ \pv{\fmfreuse{v1}}  \\
\pv{\fmfreuse{v4}}  \\ \pv{\fmfreuse{v5}}  \\
\pv{\fmfreuse{v6}}  \\ \pv{\fmfreuse{v7}}  \\
\ea \\
\eean
where all elements below those displayed
are zero (it is, of course, a square
matrix), and we have not displayed the 
zero-one-two-loop submatrices already given.
That is, the given matrix is the transpose of 
the $22 \times 6$ submatrix $B$ in 
\[
\left( 
\ba{c|c}
A & 0 \\ \hline
B & 0 \\ 
\ea \right) \; ,
\]
where $A$ is the $6 \times 6$ matrix given for $s(\pv{\fmfreuse{v1}})$
earlier. 
The order is as in table \ref{tab:graphs}, 
summarized here:
\begin{center}
\begin{tabular}{|c|c|c|c|c|c|c|c|c|c|c|}\hline 
7 & 8 & 9 & 10 & 11 & 12 & 13 & 14 & 15 & 16 & 17 \\ \hline
\rule[-3mm]{0mm}{8mm}\hspace{-1mm}
\pu{\fmfreuse{u1}}   &  \pu{\fmfreuse{u2}}  &    
\pu{\fmfreuse{u3}}   &  \pu{\fmfreuse{u4}}  &
\pu{\fmfreuse{u5}}  &  \pu{\fmfreuse{u6}} &
\pu{\fmfreuse{u7}}  &  \pu{\fmfreuse{u8}} & 
\pu{\fmfreuse{u9}}  & \pu{\fmfreuse{u10}} &
\pu{\fmfreuse{u11}} \\  \hline  \hline
18 & 19 & 20 & 21 & 22 & 23 & 24 & 25 & 26 & 27 & 28 \\ \hline
\rule[-3mm]{0mm}{8mm}\hspace{-1mm}
\pu{\fmfreuse{u12}}   &  \pu{\fmfreuse{u13}}  &    
\pu{\fmfreuse{u14}}   &  \pu{\fmfreuse{u15}}  &
\pu{\fmfreuse{u16}}  &  \pu{\fmfreuse{u17}} &
\pu{\fmfreuse{u18}}  &  \pu{\fmfreuse{u19}} & 
\pu{\fmfreuse{u20}}  & \pu{\fmfreuse{u21}} &
\pu{\fmfreuse{u22}} \\ \hline 
\end{tabular}
\end{center}
We also give two two-loop insertions:
\setlength{\arraycolsep}{1.2mm}
\bean
s^{\rm T}(\pv{\fmfreuse{v4}}){\pv{\vspace{4mm}\fmfreuse{v0}}} &=&
\left(
 \ba{cccccccccccccccccccccc}
 0 & 0 & 0 & 0 & 0 & 0 & 0 & 0 & 0 &
0 & 0 & 0 & 0 & 0 & 0 & 0 & 0 & 0 & 0 & 0 & 0 & 0  \\ \hline
 0 & 0 & 0 & 0 & \fr{2}{3} & \fr{2}{3} & \fr{2}{3} & 0 & 0 &
0 & 0 & 0 & 0 & 0 & 0 & 0 & 0 & 0 & 0 & 0 & 0 & 0  \\ \hline
 0 & 0 & 0 & 0 & 0 & 0
& 0 & 0 & 0 &
0 & 0 & 0 & 0 & 0 & 0 & 0 & 0 & 0 & 0 & 0 & 0 & 0  \\
 0 & 0 & 0 &0  & 0
& 0 & 0 & 0 & 0 &
0 & 0 & 0 & 0 & 0 & 0 & 0 & 0 & 0 & 0 & 0 & 0 & 0  \\
 0 & 0 &  0 &0 
& 0 & 0 & 0 & 0
& 0 & 0 & 0 & 0 & 0 &
0 & 0 & 0 & 0 & 0 & 0 & 0 & 0 & 0  \\
 0 & 0 & 0 & 0 & 0 & 0 & 0 & 0 & 0 &
0 & 0 & 0 & 0 & 0 & 0 & 0 & 0 & 0 &
0 & 0 & 0 & 0
\ea  \right)
\ba{c} 
\pv{\fmfreuse{v0}}  \\ \pv{\fmfreuse{v1}}  \\
\pv{\fmfreuse{v4}}  \\ \pv{\fmfreuse{v5}}  \\
\pv{\fmfreuse{v6}}  \\ \pv{\fmfreuse{v7}}  \\
\ea  \\
\eean

\bean
s^{\rm T}(\pv{\fmfreuse{v5}}){\pv{\vspace{4mm}\fmfreuse{v0}}} &=&
\left(
 \ba{cccccccccccccccccccccc}
 0 & 0 & 0 & 0 & 0 & 0 & 0 & 0 & 0 &
0 & 0 & 0 & 0 & 0 & 0 & 0 & 0 & 0 & 0 & 0 & 0 & 0  \\ \hline
 0 & 0 & 0 & \fr{1}{3} & 0 & 0 & 0 & \fr{1}{6} & \fr{1}{6} &
 \fr{1}{3} &  \fr{1}{3} &  \fr{1}{3} &  \fr{1}{3} 
& 0 & 0 & 0 & 0 & 0 & 0 & 0 & 0 & 0  \\ \hline
 0 & 0 & 0 & 0 & 0 & 0
& 0 & 0 & 0 &
0 & 0 & 0 & 0 & 0 & 0 & 0 & 0 & 0 & 0 & 0 & 0 & 0  \\
 0 & 0 & 0 &0  & 0
& 0 & 0 & 0 & 0 &
0 & 0 & 0 & 0 & 0 & 0 & 0 & 0 & 0 & 0 & 0 & 0 & 0  \\
 0 & 0 &  0 &0 
& 0 & 0 & 0 & 0
& 0 & 0 & 0 & 0 & 0 &
0 & 0 & 0 & 0 & 0 & 0 & 0 & 0 & 0  \\
 0 & 0 & 0 & 0 & 0 & 0 & 0 & 0 & 0 &
0 & 0 & 0 & 0 & 0 & 0 & 0 & 0 & 0 &
0 & 0 & 0 & 0
\ea \right) 
 \ba{c} 
\pv{\fmfreuse{v0}}  \\ \pv{\fmfreuse{v1}}  \\
\pv{\fmfreuse{v4}}  \\ \pv{\fmfreuse{v5}}  \\
\pv{\fmfreuse{v6}}  \\ \pv{\fmfreuse{v7}}  \\
\ea \\
\eean
We note that similarly to the previous case,
there is also a top left $6 \times 6$ matrix 
with entries in the first column only.
The rest of the matrices are now trivial to extract from the insertion
tables, so we shall not repeat them here.
\renewcommand{\arraystretch}{1}

\end{appendix}

\end{fmffile}

\end{document}